\documentclass[amsmath,amssymb,aps,showpacs,twocolumn,floats,prx, superscriptaddress]{revtex4-2}
\usepackage{graphicx}
\usepackage{amssymb}
\usepackage{amsmath}
\usepackage{bm}
\usepackage[colorlinks]{hyperref}
\usepackage{footnote}
\usepackage{color}
\usepackage{multirow}
\usepackage{enumerate}
\usepackage{soul}
\usepackage{lipsum}
\usepackage[normalem]{ulem}
\usepackage[FIGTOPCAP]{subfigure}
\pdfoutput=1 
\newcommand{\comm}[2]{[{#1},{#2}]}
\newcommand{\hrm}[1]{\hat{#1}^{\dagger}}

\newcommand{\expt}[1]{\left\langle #1 \right\rangle}
\newcommand{\infint}{\int^{\infty}_{\infty}}
\newcommand{\hc}{h.c.}

\newcommand{\artanh}{\mathrm{artanh}}
\newcommand{\htms}{\hat{H}_{\rm TMS}}

\begin{document}
 
\title{High-Purity Entanglement of Hot Propagating Modes Using Nonreciprocity}
 
\author{L. Orr}
\affiliation{Dahlem Center for Complex Quantum Systems and Fachbereich Physik, Freie Universit\"{a}t Berlin, 14195 Berlin, Germany}
\author{Saeed A. Khan}
\affiliation{Department of Electrical Engineering, Princeton University, Princeton, NJ 08544, USA}
\author{N. Buchholz}
\affiliation{Dahlem Center for Complex Quantum Systems and Fachbereich Physik, Freie Universit\"{a}t Berlin, 14195 Berlin, Germany}
\author{S. Kotler}
\affiliation{Racah Institute of Physics, The Hebrew University of Jerusalem, Jerusalem, 91904, Israel}
\author{A. Metelmann}
\affiliation{Dahlem Center for Complex Quantum Systems and Fachbereich Physik, Freie Universit\"{a}t Berlin, 14195 Berlin, Germany}
\affiliation{Institute for Theory of Condensed Matter, Karlsruhe Institute of Technology, 76131 Karlsruhe, Germany}
\affiliation{Institute for Quantum Materials and Technology, Karlsruhe Institute of Technology, 76344 Eggenstein-Leopoldshafen, Germany}

\date{\today}

\begin{abstract}
Distributed quantum information processing and communication protocols demand the ability to generate entanglement among propagating modes. However, thermal fluctuations can severely limit the fidelity and purity of propagating entangled states, especially for low-frequency modes relevant for radio-frequency (RF) signals. Here we propose nonreciprocity as a resource to render continuous-variable entanglement of propagating modes robust against thermal fluctuations. By utilising a cold-engineered reservoir we break the symmetry of reciprocity in a standard two-mode squeezing interaction between a low- and a high-frequency mode, and show that the rerouting of thermal fluctuations allows the generation of flying entangled states with high purity. Our approach requires only pairwise Gaussian interactions and is thus ideal for parametric circuit QED implementations.
\end{abstract}

 
\maketitle

\section{Introduction}
\label{sec_1}

Entanglement of propagating photons~\cite{horodecki_quantum_2009} is a crucial resource for quantum information processing and communication protocols~\cite{kimble_quantum_2008} and is useful for distributing entanglement amongst components of a quantum network~\cite{cirac_quantum_1997, acin_entanglement_2007}. However, as with other coherent quantum effects, it is remarkably sensitive to decoherence channels, such as thermal fluctuations. Operating at cryogenic temperatures allows for the effects of thermal fluctuations to be overcome by ensuring that $k_b T \ll \hbar \omega$, which is possible for mode frequencies $\omega$ as low as the microwave domain~\cite{eichler_observation_2012}. For even lower frequency bands, such as the radio-frequency (RF) domain that is ubiquitous in modern communication, thermal fluctuations remain appreciable even at the lowest operating temperatures~\cite{gely_observation_2019, li_stationary_2020, steele_photonics_2021}, presenting challenges for RF quantum communication and sensing. With these cryogenic temperature limitations, alternative approaches to generate entanglement in systems of hot modes must be considered.

In this paper, we present an approach utilising engineered nonreciprocity to generate steady-state entangled output fields from a system of interacting hot modes coupled to a single cold mode. While several proposals and recent experiments consider the entanglement of the output fields of cold modes coupled via an intermediate hot (for example, mechanical) mode~\cite{abdi_entangling_2015, vitali_optomechanical_2007, genes_robust_2008, wang_bipartite_2015, bienfait_phonon-mediated_2019, barzanjeh_stationary_2019, chen_entanglement_2020}, we consider situations where the fields to be entangled themselves are effectively coupled to high temperature baths. Continuous ambient thermal excitations can severely limit the entanglement fidelity of steady-state emission from such ``hot'' modes at a given pump power. Furthermore, these excitations limit the purity of the generated flying states, demanding the use of complex state purification protocols~\cite{pan_entanglement_2001,PhysRevLett.84.4002,PhysRevLett.97.150505}. We show that nonreciprocity provides a crucial ingredient to alleviate these effects: the ability to continuously reroute thermal excitations toward a cold output. This enables the entanglement of propagating photons with increased robustness to thermal excitations, and with much higher purity than is possible using a completely reciprocal two-mode entangling interaction between the hot modes of interest. 

The importance of nonreciprocity has already been firmly established in quantum information processing, enabling the routing of signals in a quantum network by realising asymmetric scattering matrices, across diverse architectures from superconducting circuits~\cite{kamal2011noiseless,sliwa2015reconfigurable,ruesink2016nonreciprocity,lecocq2017nonreciprocal,peterson2017demonstration,chapman2017widely,miri2017optical,fang2017generalized,ando2020observation} to optomechanics~\cite{shen2016experimental, bernier2017nonreciprocal, barzanjeh2017mechanical, de2019realization, xu2019nonreciprocal} and beyond~\cite{dong2015brillouin, kim2015non}. Our work analyses an aspect of nonreciprocal interactions which is much less explored: the role of nonreciprocity in manipulating \textit{fluctuations} in a quantum system, to route thermal noise while generating entanglement of the scattered fields. Building on recent progress in the theory of engineered nonreciprocity~\cite{kamal2011noiseless,Metelmann2015,ranzani2015graph,Metelmann2017}, we consider a system of three dissipative quantum modes undergoing configurable coherent interactions, and identify the conditions required for nonreciprocal scattering and directional transmission. Interestingly, by analysing the complete output state we find that entanglement can be enhanced at points of ``perfect nonreciprocity'', where scattering in one direction is forbidden. Perhaps equally as importantly, it is also possible to engineer nonreciprocal scattering between a pair of modes without entangling their outputs, highlighting the need for a deeper understanding of the connections between nonreciprocity and entanglement. To this end, we develop a heuristic picture drawing connections between steady-state entanglement in nonreciprocal systems and sequential Gaussian circuit operations, as well as dissipative entanglement schemes~\cite{wang_reservoir-engineered_2013, wang_bipartite_2015}, and ideal two-mode squeezing.

With these foundations, our work finally addresses the impact of nonreciprocity on entanglement in quantum systems experiencing thermal noise. The control of thermal noise flow~\cite{li_transforming_2021} has recently garnered renewed interest even in classical devices, due to its importance in energy harvesting, heat management, and information transfer using thermal currents~\cite{terraneo_controlling_2002, li_thermal_2004}. For quantum systems, which are our specific focus, the control of thermal noise flow becomes particularly important~\cite{joulain_quantum_2016, barzanjeh_manipulating_2018, ordonez-miranda_radiative_2019} in order to protect fragile quantum properties from thermal decoherence. To this end we show how nonreciprocal scattering can be engineered to continuously reroute incident thermal excitations away from hot modes, toward the output of the cold auxiliary mode introduced to break reciprocity. We then show that nonreciprocity can increase the entanglement fidelity and state purity of output fields scattered off the hot modes, above values that are possible using a reciprocal two-mode squeezing interaction at the same strength. Our heuristic picture shows that this increased robustness is due to a controlled swap of the input noise incident on the quantum modes at different temperatures. 

The proposed three-mode system can be efficiently realised in parametric circuit QED (cQED), where time-dependent pump fields enable tunable interactions to break reciprocity~\cite{jalas_what_2013}. Furthermore, we demand only pairwise squeezing and beam-splitter interactions, capabilities for which have been suitably demonstrated in recent cQED experiments~\cite{sliwa2015reconfigurable, lecocq2017nonreciprocal}, and which can be achieved using even a single nonlinear element based on a Josephson junction. Our model can therefore serve as a practical platform for the detailed study of quantum entanglement in the presence of nonreciprocal interactions and thermal noise.

The rest of this paper is organised as follows. In Sec.~\ref{sec:setup}, starting with a two-mode squeezing interaction, we introduce the minimal three-mode system required to render this interaction nonreciprocal, within the context of standard approaches to nonreciprocity. In Secs.~\ref{sec:entanglement} and \ref{sec:entanglementvacuum}, we proceed to analyse the scattering and entangling properties of the three-mode system, finding conditions for nonreciprocal scattering, and clarifying the connection between nonreciprocity and entanglement generation. We find that at the specific points of ``perfect nonreciprocity,'' the scattering and entangling properties of the system can be very efficiently explained as a sequence of simple pairwise linear operations. Finally, in Sec.~\ref{sec:thermal} we combine this understanding to explore the impact of thermal fluctuations on entanglement in nonreciprocal systems and demonstrate how thermal inputs to a hot mode can be efficiently routed via nonreciprocity to protect the entanglement of scattered output fields.

\section{Setup}
\label{sec:setup}

We begin with the standard description of a nondegenerate two-mode squeezing (TMS) interaction between two harmonic modes (setting $\hbar=1$),
\begin{align}
    \htms = \,\, g_{12} \left(\hrm a_1 \hrm a_2 + \hat a_1 \hat a_2 \right),
    \label{eq:HTMS}
\end{align}
where $\hat{a}_j$ is the bosonic annihilation operator for mode $j$, satisfying the standard commutation relations $[\hat{a}_j,\hat{a}^{\dagger}_k] = \delta_{jk}$. In the cQED architecture such an interaction is typically realised by appropriately pumping nonlinear Josephson-junction based superconducting elements~\cite{frattini_3-wave_2017, sivak_kerr-free_2019}.
Additionally, this interaction can be used to generate entangled photon pairs~\cite{walls_quantum_2008} and hence two-mode squeezed light for quantum information processing applications~\cite{liu_noise_2020, qiu_broadband_2022}. Nevertheless, the interaction defined by Eq.~(\ref{eq:HTMS}) is reciprocal.

\begin{figure}[tb]
		\centering
		\includegraphics[width=0.98\linewidth]{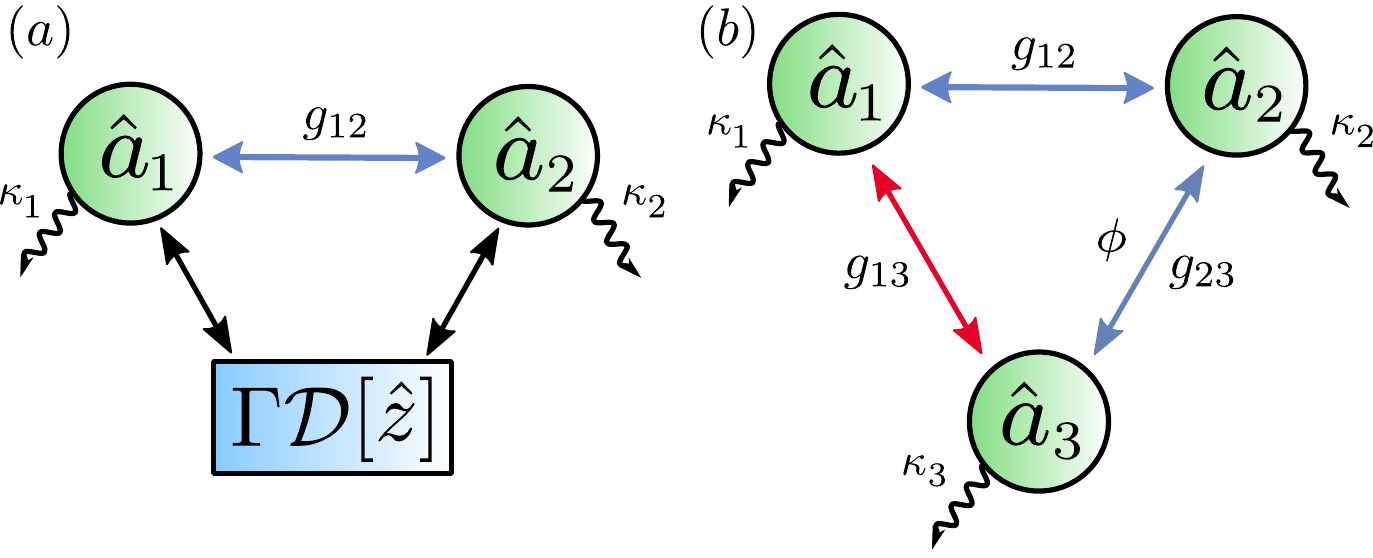}
		\caption{(a) A diagram of an open two-mode squeezer where both modes are coupled to the common nonlocal dissipator $\Gamma \mathcal{D}[\hat z]$. (b) A diagram of a minimal loop system consisting of three open modes. Modes $a_1$ and $a_2$ are coupled via a two-mode squeezing interaction, as are modes $a_2$ and $a_3$. Modes $a_1$ and $a_3$ are coupled via a beam-splitter interaction. The tunable loop phase is placed on the interaction between modes $a_2$ and $a_3$. If mode $a_3$ is adiabatically eliminated this loop is equivalent to the system depicted in (a) with the jump operator given by $\hat z = \hat a_1 + \eta e^{i \phi} \hrm a_2$.}
		\label{fig:mode_diagram}
\end{figure}

The conditions required to render interactions of the form of Eq.~(\ref{eq:HTMS}) nonreciprocal have been clarified in recent years~\cite{ranzani2015graph, Metelmann2017}. An arbitrary bidirectional interaction between two systems, $\hat{H}_{\rm int} \propto (\hat{A}\hat{B} + \hc)$ governed by operators, $\hat{A}$ and $\hat{B}$, must be balanced with a corresponding nonlocal dissipative interaction $\Gamma \mathcal{D}[\hat{z}]$ as depicted in Fig.~\ref{fig:mode_diagram}, where $\mathcal{D}[\hat{z}]\hat{\rho} = \hat{z}\hat{\rho}\hat{z}^{\dagger} - \frac{1}{2}\{\hat{z}^{\dagger}\hat{z},\hat{\rho}\}$ is the standard dissipative superoperator, with collapse operator $\hat{z} = \hat{A}+ \eta e^{i\phi}\hat{B}$. An appropriately chosen interaction strength $\Gamma$, asymmetry $\eta$, and, most crucially, phase $\phi$~\cite{Metelmann2017}, can then be used to render the desired interaction nonreciprocal.


Applying this approach to the interaction defined by Eq.~(\ref{eq:HTMS}), it is clear that dissipators with either the collapse operator $\hat{z} = \hat{a}_1 + \eta e^{i \phi} \hat{a}_2^{\dagger}$ or $\hat{z} =\hat{a}_1^{\dagger} + \eta e^{-i \phi} \hat{a}_2$ both satisfy the aforementioned form, and thus can be employed to attain the desired nonreciprocal scattering matrix. Importantly, it has been shown that either dissipative interaction alone can generate steady-state entanglement \cite{wang_reservoir-engineered_2013, wang_bipartite_2015}. Hence, by combining either dissipator with the coherent two-mode squeezing interaction in Eq.~(\ref{eq:HTMS}), one can also take the point of view that we are analysing the effects of introducing nonreciprocity in such dissipative entanglement schemes.

A hint as to what may be expected can be found in the fact that the required operators $\hat{z}$ are \textit{non-Hermitian} nonlocal collapse operators. Nonreciprocal interactions mediated by dissipators with Hermitian collapse operators have been shown to be equivalent to measurement-based feedforward schemes~\cite{Metelmann2017}: a classical observer makes a measurement on system A, and uses the result to evolve system B, breaking the reciprocity of interaction between the systems. As such, this evolution is equivalent to performing local operations and classical communication, and hence cannot generate any entanglement between the two quantum systems. In contrast, the non-Hermitian collapse operators required here have no such mapping, and can in principle generate entanglement.

To realise either nonlocal dissipator and hence render the TMS interaction nonreciprocal, we must introduce an auxiliary mode $a_3$, as depicted in Fig.~{\ref{fig:mode_diagram}(b). The open quantum system comprising this loop and the environment with which it interacts are described by the quantum optical master equation,
\begin{align}
    \dot{\hat{\rho}} = \mathcal{L}\hat{\rho} = -i[\hat{H}_{\rm NRL},\hat{\rho}] + \hspace{-1mm} \sum_{j=1,2,3} \hspace{-1.5mm} \kappa_j \mathcal{D}[\hat{a}_j]\hat{\rho},
    \label{eq:Lsys}
\end{align}
and its dynamics are governed by the Hamiltonian
	\begin{align}
		 \hat{H}_{\rm NRL}  &= \left( g_{12} \hrm a_1 \hrm a_2 
	    + g_{13} \hrm a_3 \left[\hat a_1 + \eta e^{i \phi} \hrm a_2 \right] \right) + \hc,
	    \label{eq:HNRL}
	 \end{align}
where $\eta = g_{23}/g_{13}$ accounts for an asymmetric coupling to the auxiliary mode, written in the interaction frame with respect to the three modes. Eq.~(\ref{eq:HNRL}) simply describes the original two-mode squeezing interaction between modes $a_1$ and $a_2$, but now with an auxiliary third mode that couples to mode $a_1$ via a beam-splitter interaction, and to mode $a_2$ via a two-mode squeezing interaction. When the auxiliary mode $a_3$ can be adiabatically eliminated (namely, when its damping rate $\kappa_3$ is the largest system parameter), this configuration realises the nonlocal dissipator $\hat{z} = \hat{a}_1 + \eta e^{i \phi} \hat{a}_2^{\dagger}$. Engineering of the alternative nonlocal dissipator introduced earlier leads to equivalent results. The resulting three-mode system can thus render the interaction of Eq.~(\ref{eq:HTMS}) unidirectional and can realise a nonreciprocal loop (NRL) via the general scheme described above. By explicitly including the dynamics of the auxiliary mode, we are able to explore the routing of both scattered fields and - importantly - their correlations around the loop. 

We now take a moment to discuss quantum optics platforms that can be used to realise our proposed system, and its practical implications. Eq.~(\ref{eq:HNRL}) requires only two-mode squeezing and beam-splitter interactions between pairs of modes; systems where such interactions can be controllably realised in a three-mode configuration to enable nonreciprocity have already been demonstrated in parametric cQED~\cite{sliwa2015reconfigurable, lecocq2017nonreciprocal}. For concreteness, in Appendix~\ref{app:parCQED}, we detail how a simple three-mode circuit with a single nonlinear mixing element can realise the NRL. The required pairwise interactions have also been realised across other platforms such as optomechanical circuits~\cite{peterson_demonstration_2017, bernier_nonreciprocal_2017, fang_generalized_2017}. Importantly, the three modes constituting the NRL can have widely distinct frequencies, and thus experience thermal effects to varying degrees (see Appendix~\ref{app:parCQED}). The coupling of modes with such disparate frequencies has been realised in very recent experiments~\cite{rodrigues_cooling_2021, rodrigues_parametrically_2022}, and is highly relevant to RF-domain quantum optics. In precisely these cases, the NRL provides a way of enhancing the entanglement and purity of hot propagating modes using auxiliary cold modes, which is a key result of our work.

\section{Scattering properties and nonreciprocity }
\label{sec:entanglement}

When analysing the three-mode loop, we restrict the initial state $\hat \rho_{\mathrm{in}}$ of the system to Gaussian states. Each mode is equipped with a pair of quadrature operators, $\hat X_j = (\hrm a_j + \hat a_j)/\sqrt{2}$ and $\hat P_j = i (\hrm a_j - \hat a_j)/\sqrt{2}$ where $j\in\{1,2,3\}$. We can also define quadrature modes for the input or output states by replacing $j$ with $(j,\{\mathrm{in},\mathrm{out}\})$, respectively. By declaring the initial state $\hat \rho_{\mathrm{in}}$ to be Gaussian, we mean that it is completely characterised by the first and second moments of these quadrature operators. Since the dynamics induced by the Hamiltonian in Eq.~(\ref{eq:HNRL}) are entirely linear, the state of the system at any time will remain Gaussian, including the output state $\hat\rho_{\mathrm{out}}$~\cite{weedbrook2012gaussian}.

Our approach to analysing the steady-state scattering properties of the system is standard: we solve the linear Heisenberg-Langevin equations in frequency space, and use quantum input-output theory~\cite{gardiner_input_1985} (see Appendix~\ref{app:heis}). From this, we obtain the scattering matrix $\mathbf{S}[\omega]$ relating output-field operators to the input fields,
	\begin{equation}
	\vec R_{\mathrm{out}}[\omega] = \mathbf{S}[\omega] \vec R_{\mathrm{in}}[\omega],
	\label{eq:smatdef}
	\end{equation}
where $\vec{R}_{\{\mathrm{in},\mathrm{out}\}}[\omega]$ is a vector of quadrature operators for the input and output modes, respectively.

We are interested in the scattering behaviour on resonance, that is, when $\omega=0$ in this frame, which describes the response of fields resonant with the individual modes comprising the system. The scattering matrix in this simpler case, $\mathbf{S}[0] \equiv \mathbf{S}$, can be expressed purely in terms of the cooperativities $\mathcal{C}_{jk} = 4 g_{jk}^2 / \kappa_j \kappa_k$ parameterising interactions between modes $a_j$ and $a_k$ in relation to their individual decay rates. By balancing the cooperativities and adjusting the loop phase $\phi$, scattering between any pair of modes in the system can be rendered nonreciprocal. The full form of the scattering matrix is still unwieldy (see Appendix~\ref{app:smatblock}). We therefore introduce measures that allow us to quantify the scattering properties of the system more compactly.

For nonreciprocal systems, we are primarily interested in the asymmetry of scattering between modes $j$ and $k$. To more precisely quantify this asymmetry, we introduce the normalised degree of nonreciprocity $\mathcal{N}^{(j,k)}$ as
\begin{align}
    \mathcal{N}^{(j,k)} = \frac{||{\rm abs}~\mathbf{S}_{jk}-{\rm abs}~\mathbf{S}_{kj}||}{||\mathbf{S}_{jk}||+||\mathbf{S}_{kj}||}
    \label{eq:nij}
\end{align}
where $\mathbf{S}_{jk}$ is the two-mode block of the scattering matrix corresponding to modes $j$ and $k$, ${\rm abs}~\mathbf{O}$ is an element-wise absolute value operation, and $|| \cdot ||$ is the Frobenius norm. $\mathcal{N}^{(j,k)}$ is a quantity that remains bounded within $[0,1]$, and measures the difference in \textit{amplitude} (and not phase) of scattering between a pair of modes. As a result, it vanishes for scattering that is reciprocal in amplitude but that may differ in phase. As discussed in Appendix~\ref{app:smatblock}, it can be shown that $\mathcal{N}^{(j,k)} = 0~\forall~j\neq k$ only when $\phi = 0$.

For all other values of $\phi$, the three-mode system exhibits nonreciprocal scattering properties to varying degrees. To understand how this nonreciprocity influences the quantum properties of scattered input fields, we can analyse the covariance matrix of the output fields. However, before analysing the general case, we find that several key ideas can be understood via a simple heuristic picture that is valid when $\mathcal{N}^{(j,k)} = 1$. From Eq.~(\ref{eq:nij}), this corresponds to perfectly asymmetric scattering, where either $||\mathbf{S}_{jk}||$ or $||\mathbf{S}_{kj}||$ vanishes. We refer to these as points of \textit{perfect} nonreciprocity and the required conditions are summarised below. The arrows denote the direction in which signal transmission is allowed; scattering matrix elements in the reverse direction vanish exactly:
	\begin{equation}
	\begin{array}{|c||c|c|}
	\hline
	\hspace{1mm} & \phi = -\pi/2 & \phi = +\pi/2 \\ \hline \hline
	\mathbf{\mathcal{C}_{12} = \mathcal{C}_{13} \mathcal{C}_{23}} & \bm{a_1 \rightarrow a_2} & \bm{a_1 \leftarrow a_2} \\ \hline
    \mathcal{C}_{23} = \mathcal{C}_{12} \mathcal{C}_{13} & a_2 \rightarrow a_3 & a_2 \leftarrow a_3 \\ \hline
    \mathcal{C}_{13} = \mathcal{C}_{12} \mathcal{C}_{23} & a_1 \rightarrow a_3 & a_1 \leftarrow a_3 \\ \hline 
	\end{array}.
	\label{eq:NRCond}
	\end{equation}
In this work, we pay particular attention to scattering between modes $a_1$ and $a_2$, and hence set $\mathcal{C}_{12} = \mathcal{C}_{13} \mathcal{C}_{23}$. Plotting $\mathcal{N}^{(1,2)}$ in Fig.~{\ref{fig:circuit_diagram}}(a), we see that scattering between modes $a_1$ and $a_2$ can be rendered perfectly nonreciprocal, $\mathcal{N}^{(1,2)} = 1$, when $\phi = \pm \pi/2$. 

\subsection{Circuit decomposition at points of perfect nonreciprocal scattering}

The scattering matrix describes a potentially complicated set of linear operations on Gaussian states. In order to more easily understand the scattering behaviour of the system, a variety of decomposition schemes can be used to represent this set of operations more efficiently. One such prominent example is the Bloch-Messiah decomposition~\cite{braunstein2005squeezing}, a special case of the singular value decomposition for the group of real symplectic matrices, of which the steady-state scattering matrix $\mathbf{S}$ is an element, $\mathbf{S} \in Sp(2n,\mathbb{R})$.

We find that for the perfectly nonreciprocal system under consideration, the less commonly used \textit{polar} decomposition~\cite{gosson_symplectic_2011} proves to be simpler. This decomposition allows us to write a symplectic matrix in the form $\mathbf{RU}$, referred to as its left polar decomposition, where $\mathbf{U} \in Sp(2n,\mathbb{R}) \cap O(2n,\mathbb{R})$ is a real symplectic orthogonal matrix, and $\mathbf{R} \in Sp(2n,\mathbb{R}) \cap \mathrm{Sym}^+(2n)$ is a real symplectic symmetric positive definite matrix. Physically, the matrix $\mathbf{U}$ represents passive optical transformations, namely beam-splitters and phase shifters. The matrix $\mathbf{R}$ then includes any single and two-mode squeezing interactions. An equivalent form $\mathbf{UR}$ of the scattering matrix is provided by the right polar decomposition, where $\mathbf{R}$ and $\mathbf{U}$ are in general distinct from the left polar decomposition.

In general, the polar decomposition leads to dense and complicated matrices $\mathbf{R}$ and $\mathbf{U}$. Remarkably, we find that the scattering matrix describing our three-mode system has extremely simple forms for the left and/or right polar decomposition at points of perfect nonreciprocal scattering, $\mathcal{N}^{(1,2)}=1$, up to a global change in phase on the scattering matrix, $-\mathbf{S}$. Due to the structure of the covariance matrix [see Eq.~(\ref{eq:VW})] this change of phase will not affect the resulting covariances. In these simple cases, the $\mathbf{R}$ and $\mathbf{U}$ matrices involve only a single interaction between one pair of modes. The polar decomposition when $\mathcal{N}^{(1,2)}=1$ is then comprised of symplectic matrices corresponding to the following two unitary operations:
    \begin{align}\label{eq:PolarDecomp}
    	\mathbf{U}^{(1,3)} \leftrightarrow & \exp \left[2 i \arctan 
    	\Big(\!\sqrt{\mathcal{C}_{13}}\Big) \left(\hrm a_1 \hat a_3 + \hat a_1 \hrm a_3 \right) \right] \nonumber \\
	    \mathbf{R}^{(2,3)} \leftrightarrow & \exp \left[- 2 \, \artanh \Big(\!\sqrt{\mathcal{C}_{23}}\Big) \left(\hrm a_2 \hrm a_3 - \hat a_2 \hat a_3 \right) \right].
	\end{align}
Crucially, we find that both left and right polar decompositions can provide useful complementary insights into the action of the three-mode system at points of perfect nonreciprocal scattering.

\begin{figure}[t]
		\centering
		\includegraphics[width=1.0\linewidth]{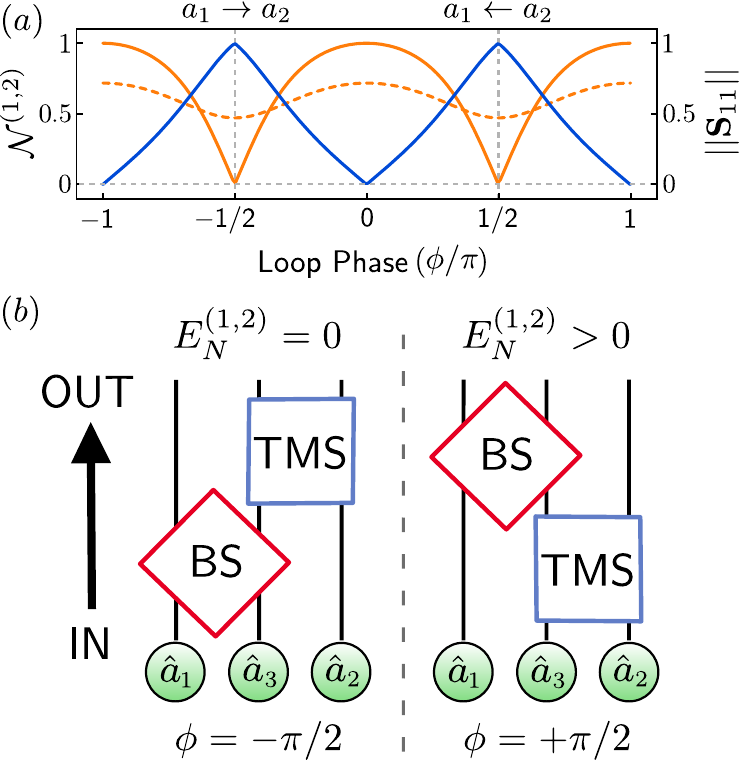}
		\caption{(a) The scattering properties of the nonreciprocal loop as a function of the loop phase $\phi$. $\mathcal{N}^{(1,2)}$ is plotted in blue for the impedance-matched case only; the general case is qualitatively similar and is omitted for clarity. $||\mathbf{S}_{11}||$ is plotted in orange for the impedance-matched (solid) and general (dashed) cases. (b) Circuit descriptions of the nonreciprocal system at points of perfect nonreciprocal scattering, where $\mathcal{N}^{(1,2)} = 1$. The form of these circuits holds regardless of the value of $||\mathbf{S}_{11}||$. The left diagram corresponds to $\phi = -\pi/2$ which yields perfect nonreciprocal scattering in the $a_1\rightarrow a_2$ direction, while the right diagram corresponds to $\phi = \pi/2$ where scattering is only allowed in the $a_1\leftarrow a_2$ direction. The input is initially uncorrelated and moves from bottom to top.}
		\label{fig:circuit_diagram}
\end{figure}

For the loop phase $\phi = - \pi/2$, it is the left polar decomposition that takes on a simple form:
 	\begin{equation}\label{eq:PDMinusPhi}
 	- \mathbf{S} = \mathbf{R}^{(2,3)} \, \mathbf{U}^{(1,3)} \text{ when } \phi = - \frac{\pi}{2}.
 	\end{equation}
As illustrated in Fig.~\ref{fig:circuit_diagram}, modes $a_1$ and $a_3$ interact first via a beam-splitter $\mathbf{U}^{(1,3)}$ which acts to exchange input from $a_1$ to $a_3$, and vice versa. This operation is followed by a two-mode squeezer $\mathbf{R}^{(2,3)}$ between modes $a_2$ and $a_3$, where the output of mode $a_2$ then becomes dependent on the input of mode $a_1$, thus realising directional transmission from $a_1 \rightarrow a_2$. The output of mode $a_1$ cannot have any dependence on the input of $a_2$ because it can only arrive at the output of mode $a_3$ via the same two-mode squeezer.

For the opposite sign of the loop phase, $\phi = +\pi/2$, the right polar decomposition yields
    \begin{equation}\label{eq:PDPlusPhi}
 	- \mathbf{S} = \mathbf{U}^{(1,3)} \, \mathbf{R}^{(2,3)} \text{ when } \phi = \frac{\pi}{2},
    \end{equation}
which describes the same component operations as Eq.~(\ref{eq:PDMinusPhi}) but applied in \textit{reverse} order. As a result, the scattering behaviour is reversed and still nonreciprocal, allowing transmission from $a_1 \leftarrow a_2$. 
 
The relatively simple form of these circuits provides a heuristic picture of nonreciprocal scattering in this system, where changing the direction of the nonreciprocal scattering is equivalent to changing the order of operations in the circuit. However, note that the nonreciprocal behaviour is not explained just by the sequential beam-splitters, but also involves two-mode squeezing interactions. This already hints at the possibility of generating nontrivial quantum correlations in scattered output fields and thus connects to entangling properties of the three-mode system, as we will see in Sec.~\ref{sec:entanglementvacuum}.

\section{Output entanglement and purity}
\label{sec:entanglementvacuum}

The output state of the three-mode system, and all of its entangling capabilities, is completely characterised by the generally frequency-dependent covariance matrix of the output-field quadrature operators, $\mathbf{V}[\omega]$, which can be calculated using Eq.~(\ref{eq:smatdef}) and the known correlation relations of the input-field operators. The output covariance matrix can then be written in terms of the scattering matrix
 	\begin{equation}
 	\mathbf{V}[\omega] = \frac{1}{2} \left(\mathbf{S}[\omega] \mathbf{V}_{\mathrm{in}} \mathbf{S}^T[-\omega] + \mathbf{S}[-\omega] \mathbf{V}_{\mathrm{in}} \mathbf{S}^T[\omega] \right),
 	\label{eq:VW}
 	\end{equation}
where $\mathbf{V}_{\mathrm{in}}$ is the matrix of correlations of the input fields. Assuming that the input noise for different modes is uncorrelated, $\mathbf{V}_{\mathrm{in}}$ contains variances determined by the thermal occupation number $n_j^{\mathrm{th}}$ and vacuum fluctuations (see Appendix \ref{app:heis}),
	\begin{equation}\label{eq:v0def}
	\mathbf{V}_{\mathrm{in}} = \bigoplus_{j=1}^{n} \left(n_j^{\mathrm{th}} + \frac{1}{2}\right) \mathbf{I}
	\end{equation} 
where $n=3$ is the number of modes and $\mathbf{I}$ is the $2 \times 2$ identity matrix. Once again considering the response on resonance, we set $\omega=0$ in Eq.~(\ref{eq:VW}) so that the covariance matrix of interest, $\mathbf{V}[0] \equiv \mathbf{V}$, takes the simple form
\begin{equation}
        \mathbf{V} =  \mathbf{S} \mathbf{V}_{\mathrm{in}} \mathbf{S}^T.
        \label{eq:V0}
\end{equation}
From this covariance matrix, we aim to calculate useful entanglement metrics for different bipartitions of the output fields in order to investigate the effects of nonreciprocal scattering. These include the Simon-Peres-Horodecki criterion for the separability of Gaussian states~\cite{simon2000peres}, as well as the logarithmic negativity $E_{N}^{(j,k)}$, an entanglement monotone for the shared output state of modes $a_j$ and $a_k$~\cite{vidal2002computable,plenio2005logarithmic,weedbrook2012gaussian} for the shared output state of modes $a_j$ and $a_k$. The latter may be calculated from the minimum symplectic eigenvalue $\nu_-^{(j,k)}$ of the partial transpose of the corresponding two-mode block $\mathbf{V}^{(j,k)}$ from the total covariance matrix $\mathbf{V}$ via
	\begin{equation}
	E^{(j,k)}_N = \begin{cases} 0 & \quad \text{for } 2 \nu_-^{(j,k)} \geq 1 \\ -\log \left(2 \nu_-^{(j,k)}\right) & \quad \text{for } 2 \nu_-^{(j,k)} < 1. \end{cases}
	\end{equation}
Similarly, we can define the marginal purity $\mu^{(j,k)}$ of a given bipartition of the output field of modes using
	\begin{equation}\label{eqn:purity}
	\mu^{(j,k)} = \frac{1}{4 \sqrt{\det \mathbf{V}^{(j,k)}}}.
	\end{equation}
This measure has a maximum value of $\mu^{(j,k)}=1$ only for pure states; for mixed states the purity will be $\mu^{(j,k)}<1$.

\subsection{Entanglement and purity in a nonreciprocal system}

We begin by examining the entanglement properties and purity of output fields under vacuum input, $n^{\mathrm{th}}_j = 0$, for all modes. In this instance the initial matrix of correlations is comprised of vacuum noise and is therefore the identity matrix, $\mathbf{V}_{\mathrm{in}} = \mathbf{I}_6/2$, so the output covariance matrix is $\mathbf{V} = \frac{1}{2}\mathbf{S} \mathbf{S}^T$. Balancing the cooperativities $\mathcal{C}_{12} = \mathcal{C}_{13} \mathcal{C}_{23}$, we plot the logarithmic negativity between the output of modes $a_1$ and $a_2$, $E_N^{(1,2)}$, as well as $E_N^{(2,3)}$, in Fig.~\ref{fig:vacuum_ent_purity}(a), where we note the strong dependence on the value of the loop phase. The simple form for the circuit decomposition means that the behaviour of the entanglement at the points of perfect nonreciprocal scattering between modes $a_1$ and $a_2$, corresponding to the points where $\phi = \pm\pi/2$ in Fig.~\ref{fig:vacuum_ent_purity}, can be explained simply as follows.

The covariance matrix for the $a_1 \rightarrow a_2$ direction of perfect nonreciprocal scattering (where $\phi=-\pi/2$) may be written using the form of the scattering matrix from Eq.~(\ref{eq:PDMinusPhi}):
\begin{equation}\label{eq:V0Minus}
        \mathbf{V} = \frac{1}{2} \mathbf{R}^{(2,3)} \big(\mathbf{R}^{(2,3)}\big)^T
\end{equation}
where the beam-splitter component does not appear because it is an orthogonal transformation. The entangling behaviour for this direction of the nonreciprocal scattering is therefore equivalent to a two-mode squeezer between modes $a_2$ and $a_3$; hence we must have $E^{(2,3)}_N>0$. Since modes $a_1$ and $a_2$ do not share any squeezing in this representation, there will be no entanglement generated between these two modes, which is evident in Fig.~\ref{fig:vacuum_ent_purity}.

We can then ask whether there are other operating points where the entanglement between the output of modes $a_1$ and $a_2$ vanishes. It is possible to determine this for all system parameters, and not just at the points of perfect nonreciprocity, by examining the Simon-Peres-Horodecki criterion for the output of modes $a_1$ and $a_2$:
\begin{equation}\label{eqn:Simon_Crit_Ent12}
    \sqrt{\frac{\mathcal{C}_{12}}{\mathcal{C}_{13} \mathcal{C}_{23}}} + \sqrt{\frac{\mathcal{C}_{13} \mathcal{C}_{23}}{\mathcal{C}_{12}}} \leq -2 \sin \phi.
\end{equation}
The output fields of modes $a_1$ and $a_2$ are separable so long as the above inequality is satisfied, which only occurs for one set of parameters: $\mathcal{C}_{12} = \mathcal{C}_{13} \mathcal{C}_{23}$ and $\phi = - \pi/2$, which is the point of perfect nonreciprocal scattering where $a_1 \rightarrow a_2$. This is, then, the only point of operation where the entanglement of the output of modes $a_1$ and $a_2$ vanishes: $E^{(1,2)}_N=0$. Away from this point, the output for these two modes will always be entangled.

For the reverse direction, $a_1 \leftarrow a_2$ (where $\phi = +\pi/2$), we can use Eq.~(\ref{eq:PDPlusPhi}) to write the covariance matrix as
\begin{equation}\label{eq:V0Plus}
    \mathbf{V} = \frac{1}{2} \mathbf{U}^{(1,3)} \mathbf{R}^{(2,3)} \big(\mathbf{R}^{(2,3)}\big)^T \! \big(\mathbf{U}^{(1,3)}\big)^T.
\end{equation}
Importantly, the beam-splitter between modes $a_1$ and $a_3$ appears and therefore plays an important role in the entanglement generation here. The two-mode squeezer acts first to entangle modes $a_2$ and $a_3$, while the later action of the beam-splitter swaps some of these squeezed correlations from mode $a_3$ to $a_1$, generating entanglement between the output of modes $a_1$ and $a_2$. This resembles the dissipative entanglement protocol \cite{wang_reservoir-engineered_2013, wang_bipartite_2015}, where the entanglement of two modes is mediated by a strongly damped auxiliary mode.

Moreover, at $\phi = \pi/2$, there exists only one specific mode of operation where the additional entanglement between mode $a_2$ and $a_3$ vanishes: when $\mathcal{C}_{13}=1$, the beam-splitter in Eq.~(\ref{eq:PolarDecomp}) acts to \textit{perfectly} swap all squeezing from mode $a_3$ to $a_1$ and swap all the uncorrelated vacuum noise from mode $a_1$ to $a_3$. The result of this perfect swap is that $a_1$ and $a_2$ now form a two-mode squeezed vacuum state, so the value of $E_N^{(1,2)}$ will be equivalent to the value achieved by a TMS, as seen in Fig.~\ref{fig:vacuum_ent_purity}. Since modes $a_2$ and $a_3$ no longer share any squeezed correlations their measure of entanglement must vanish: $E_N^{(2,3)} = 0$.

\begin{figure}[t]
		\centering
		\includegraphics[width=0.9\linewidth]{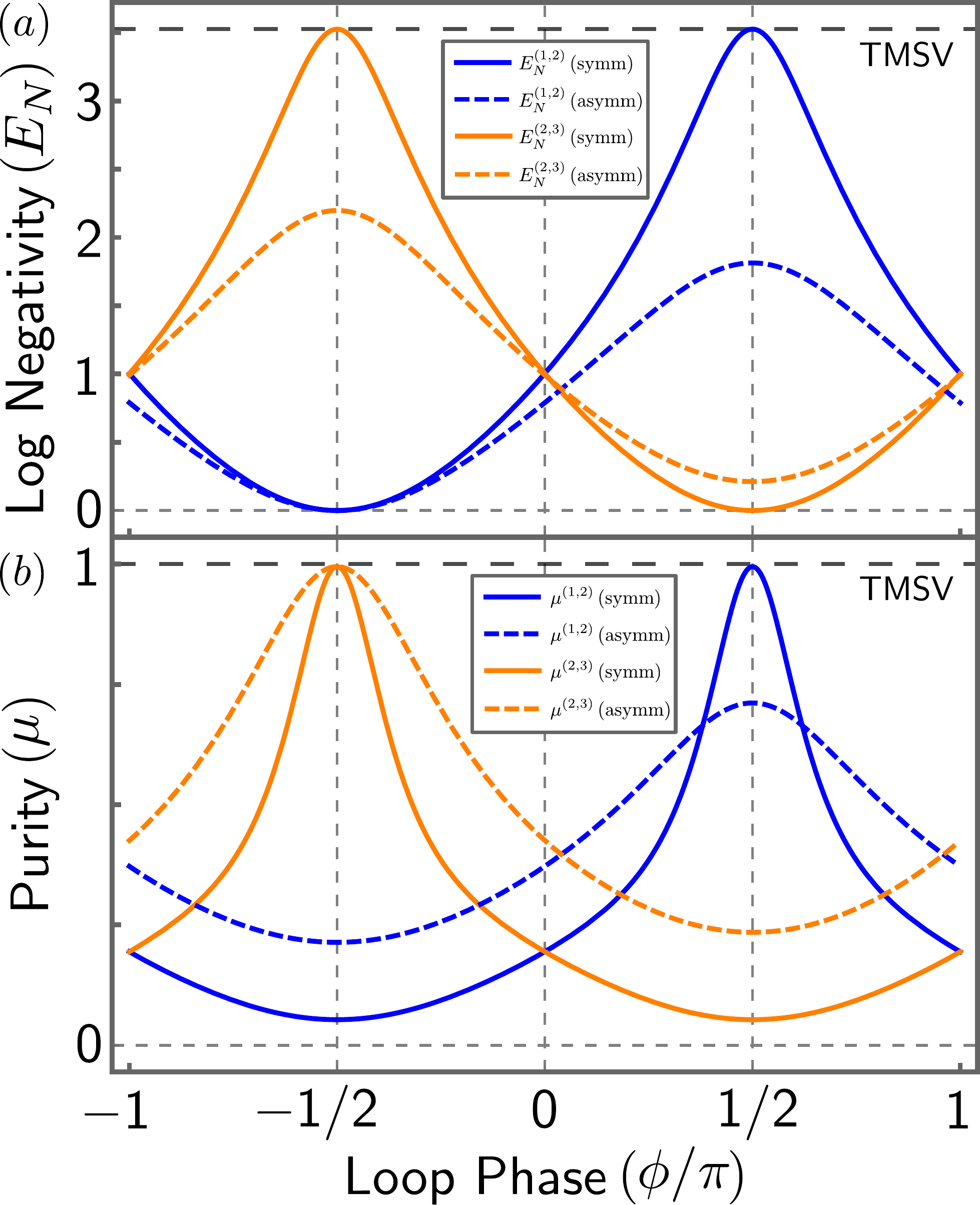}
		\caption{(a) The logarithmic negativity $E_N^{(j,k)}$ and (b) the purity $\mu^{(j,k)}$ of the stationary output states when $\mathcal{C}_{12} = \mathcal{C}_{13} \mathcal{C}_{23}$ as a function of the loop phase. The results for the joint states of the outputs of modes $a_1$ and $a_2$ (blue) and $a_2$ and $a_3$ (orange) are shown, for two parameter choices: ``symmetric,'' $\mathcal{C}_{23}=\mathcal{C}_{12},\mathcal{C}_{13} = 1$ (solid) and ``asymmetric,'' $\mathcal{C}_{23}=\mathcal{C}_{12}/2,\mathcal{C}_{13} = 2$ (dashed), with $\mathcal{C}_{12} = 0.5$ in both cases. All baths are in the vacuum state. The dashed black lines at the top of each plot give results for the output state of a \textit{reciprocal} two-mode squeezed system with vacuum input (TMSV), $\mathcal{C}_{13}=\mathcal{C}_{23}=0$.}
		\label{fig:vacuum_ent_purity}
\end{figure}

In order to discuss the behaviour of $E^{(2,3)}_N$ in more detail we again use the Simon-Peres-Horodecki criterion; the output of $a_2$ and $a_3$ will be separable so long as the following inequality is satisfied:
\begin{equation}\label{eqn:Simon_Crit_Ent23}
    \sqrt{\frac{\mathcal{C}_{23}}{\mathcal{C}_{12} \mathcal{C}_{13}}} + \sqrt{\frac{\mathcal{C}_{12} \mathcal{C}_{13}}{\mathcal{C}_{23}}}\leq 2 \sin \phi.
\end{equation}
Referring to Eq.~(\ref{eq:NRCond}), it is evident that this is only satisfied when there is nonreciprocal scattering with direction $a_2 \leftarrow a_3$. The required phase here is $\phi=+\pi/2$ which is the \textit{opposite} phase requirement from Eq.~(\ref{eqn:Simon_Crit_Ent12}).

It is evident that the degree and direction of nonreciprocal scattering, and therefore the value of the phase $\phi$, plays a crucial role in the behaviour of the output-field entanglement, as depicted in Fig.~\ref{fig:vacuum_ent_purity}. In particular, when $\mathcal{C}_{12}=\mathcal{C}_{23}$ and $\mathcal{C}_{13} =1$, $\mathcal{N}^{(1,2)}=1$ and $\mathcal{N}^{(2,3)}=1$, so the scattering processes for both pairs of modes are perfectly nonreciprocal. $E_{N}^{(1,2)}$ and $E_{N}^{(2,3)}$ reach maximum values for this parameter regime, where they can both realise the same entangling power of a reciprocal two-mode squeezer; however, the maxima are achieved at different values of the phase. In addition this is also the only operating regime where both $E_N^{(1,2)}$ and $E_N^{(2,3)}$ reach the absolute minimum value of 0.

Furthermore, in this parameter regime, where perfect swapping is also observed, the scattering of modes $a_1$ and $a_3$ is impedance matched in both cases ~(see Appendix~\ref{app:smatblock}), so the input noise is not reflected in the output fields. For later convenience, we refer to the regime where $\mathcal{C}_{13}=1$ as the ``symmetric'' case, since $\mathcal{C}_{12}=\mathcal{C}_{23}$. We therefore label the regime where $\mathcal{C}_{13} \neq 1$ the ``asymmetric'' case. The degree of impedance on mode $a_1$ is presented in Fig.~\ref{fig:circuit_diagram}(a), where we see that the reflection of mode $a_{1}$ vanishes, i.e., $|| \mathbf{S}_{11} || = 0$, only in the symmetric case.

The circuit decomposition also allows for a heuristic explanation of the behaviour of the marginal purities, seen in Fig.~\ref{fig:vacuum_ent_purity}(b). Since the initial covariance matrix is $\mathbf{V}_{\mathrm{in}} = \mathbf{I}_6/2$ for vacuum inputs, the marginal purities for the input states will be $\mu^{(1,2)}=1$ and $\mu^{(2,3)}=1$. These purities will remain unchanged in the output state provided that the corresponding two-mode block of the output covariance matrix can be reached by a symplectic transformation. This follows since the determinant of a symplectic transformation is $\mathrm{det}(\mathbf{S})=1$ and therefore $\mathrm{det}(\mathbf{S} \mathbf{O} \mathbf{S}^T) = \mathrm{det}(\mathbf{O})$ for any matrix $\mathbf{O}$. 

Since the covariance matrix for $\phi = - \pi/2$ [see Eq.~(\ref{eq:V0Minus})] simply describes a two-mode squeezing interaction between modes $a_2$ and $a_3$, the marginal purity for their outputs will always remain the same, $\mu^{(2,3)} = 1$.. On the other hand, the marginal purity of the output state between modes $a_1$ and $a_2$ is below one, $\mu^{(1,2)}<1$, in this case, as the two-mode block $\mathbf{V}^{(1,2)}$ cannot be reached by any combination of symplectic transformations. For the opposite phase $\phi = +\pi/2$, this same reasoning applies to the two-mode block $\mathbf{V}^{(2,3)}$, so that now $\mu^{(2,3)} < 1$. However, $\mu^{(1,2)}$ is still generally less than one, with a notable exception: the symmetric case, where $\mathbf{V}^{(1,2)}$ becomes equivalent to the covariance matrix for a two-mode squeezed vacuum state. In fact, the condition that saturates the marginal purity $\mu^{(1,2)}=1$ is the same condition for which only the output of modes $a_1$ and $a_2$ are entangled, as $E_N^{(2,3)} = 0$ [see Eq.~(\ref{eqn:Simon_Crit_Ent23})].

We have therefore observed and explained how entanglement arises in a nonreciprocal system. Crucially, we find that the points of perfect nonreciprocity play a special role in maximising the achievable output entanglement. We also note how the special symmetric case allows for the purity of the output states to be maximised. We are now in a position to analyse the role of nonreciprocity in entanglement generation in the presence of thermal fluctuations.

\section{Output entanglement in the presence of thermal noise}
\label{sec:thermal}

\begin{figure}[t]
		\centering
		\includegraphics[width=0.9\linewidth]{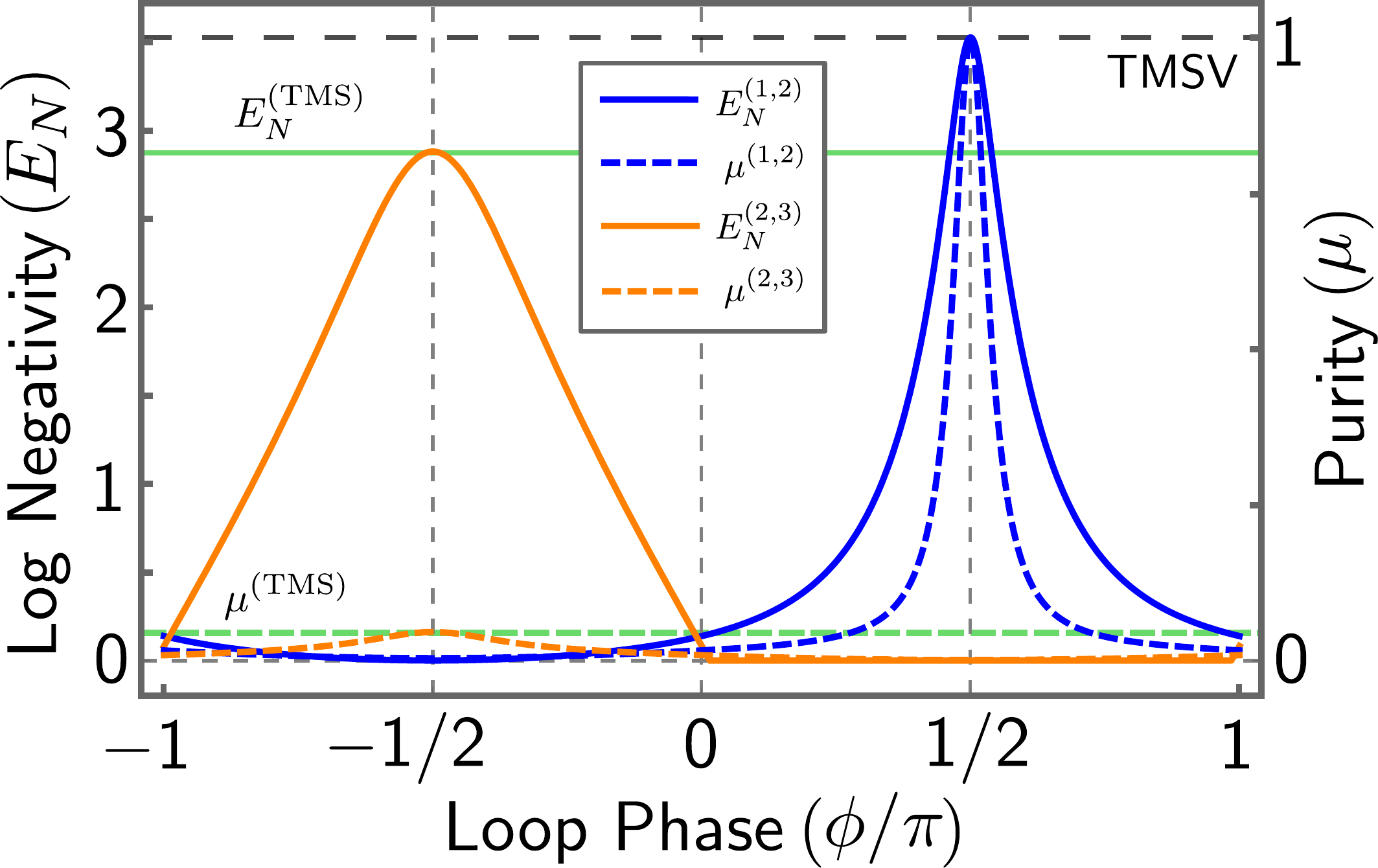}
		\caption{The logarithmic negativity (solid) and purity (dashed) of NRL output states as a function of the loop phase in the presence of thermal noise. Here, $n^{\rm th}_1=10$ and $n^{\rm th}_2=n^{\rm th}_3=0$. The results shown are for the symmetric case where $\mathcal{C}_{12}=\mathcal{C}_{23} = 0.5$ and $\mathcal{C}_{13} = 1$, describing the joint output states of modes $a_1$ and $a_2$ (blue), and modes $a_2$ and $a_3$ (orange). We also show the results for a two-mode squeezed state where $n_1^{\mathrm{th}}=10$ and $n_2^{\mathrm{th}}=0$ (green). The dashed black line indicates both the logarithmic negativity and purity for a two-mode squeezed vacuum state (TMSV).}
		\label{fig:thermal_ent_purity_phase}
\end{figure}

Thermal noise is an unwanted feature when attempting to generate entanglement. In the case of a reciprocal two-mode squeezer, thermal noise incident on one or both modes will only serve to degrade the logarithmic negativity. While this can be overcome by increasing the strength of the two-mode squeezing interaction (e.g., in parametric cQED, using stronger pump strengths), the same is not true for the purity of the generated output state. More precisely, the purity for a two-mode squeezed system where the thermal noise at the inputs for both modes is $n_1^{\mathrm{th}}$ and $n_2^{\mathrm{th}}$ is given by $\mu^{(1,2)} = 1/(2 n_1^{\mathrm{th}}+1)(2 n_2^{\mathrm{th}}+1)$, which is independent of the degree of squeezing.

One might expect that reciprocally coupling an auxiliary cold mode to the hot modes of interest would help in mitigating this impact. However, while such a coupling can reduce the internal occupation of the hot modes, it is unable to continuously route thermal inputs in a specified direction: away from the propagating output fields. Combining this cold auxiliary mode with nonreciprocity enables unidirectional scattering of coherent input signals which extends to the routing of thermal fluctuations, while also allowing for the output of the target modes to be entangled. While it is also possible for a three-mode reciprocal system to route thermal fluctuations in a similar manner, no entanglement can be generated between the outputs of the target modes (see Appendix~\ref{sec:reciprocalrerouting}).

\begin{figure}[t]
		\centering
		\includegraphics[width=0.9\linewidth]{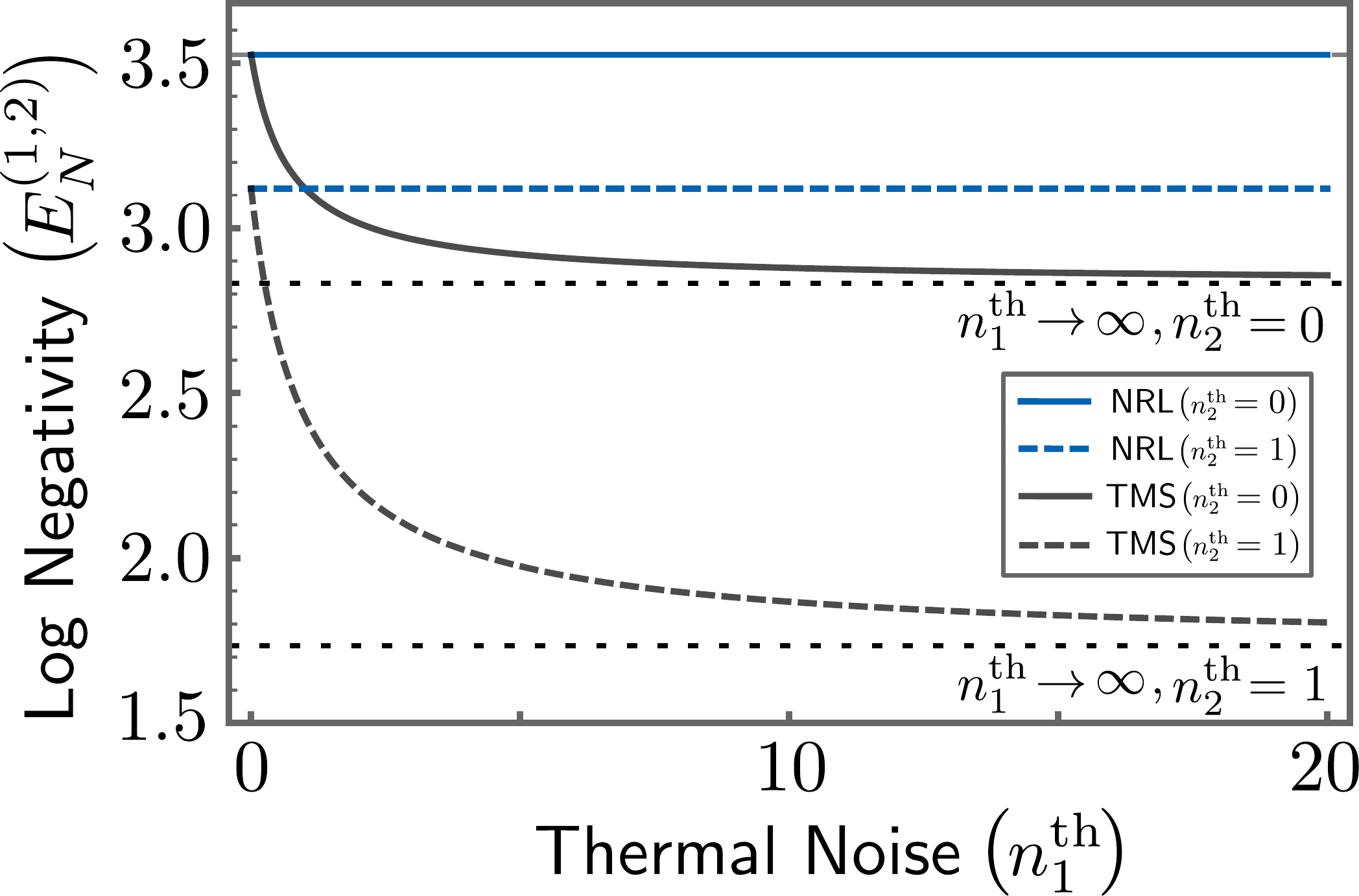}
		\caption{The entanglement between the stationary output of the modes $a_1$ and $a_2$ as measured by the logarithmic negativity, as a function of the strength of the thermal noise input $n_1^{\rm th}$ on mode $a_1$. We compare the results from an open two-mode squeezed system (TMS, grey) with our nonreciprocal three-mode loop (NRL, blue). In both cases, the squeezing interaction between modes $a_1$ and $a_2$ has cooperativity $\mathcal{C}_{12} = 0.5$. The symmetric case is shown for the nonreciprocal loop, so $\mathcal{C}_{23} = 0.5$, $\mathcal{C}_{13} = 1$,  and $\phi = +\pi/2$. The thermal occupation of mode $a_2$ is taken to be $n^{\rm th}_2=0$ (solid) and $n^{\rm th}_2=1$ (dashed). Mode $a_3$ is always taken to have vacuum input. The dashed black lines correspond to the logarithmic negativity for a TMS where $n_1^{\rm th} \rightarrow \infty, n_2^{\rm th} = 0$ and $n_1^{\rm th} \rightarrow \infty, n_2^{\rm th} = 1$.}
		\label{fig:thermal_variation}
\end{figure}

\subsection{Rerouting thermal fluctuations using nonreciprocity}

The nonreciprocal loop provides a way for us to avoid these effects on the state shared between the output of modes $a_1$ and $a_2$ by setting the parameters to the symmetric case, where the scattering of modes $a_1$ and $a_3$ are both impedance matched. Provided that the input of mode $a_3$ is vacuum noise, if we set $\phi = +\pi/2$, thermal noise in the input of mode $a_1$ can be rerouted to the output of mode $a_3$. Due to the presence of thermal noise, the initial covariance matrix is no longer proportional to the identity matrix and so the output covariance matrix has the following form:
\begin{equation}\label{eq:VThPlus1}
    \mathbf{V} = \mathbf{U}^{(1,3)}_{\mathrm{swap}} \mathbf{R}^{(2,3)} \mathbf{V}_{\mathrm{in}} \big(\mathbf{R}^{(2,3)}\big)^T \! \big(\mathbf{U}^{(1,3)}_{\mathrm{swap}}\big)^T,
\end{equation}
where $\mathbf{U}_{\rm swap}^{(1,3)}$ is the beam-splitter operation from Eq.~(\ref{eq:PolarDecomp}) when $\mathcal{C}_{13} =1$, which describes a perfect swap. However, the circuit description shown in Fig.~\ref{fig:circuit_diagram} still holds: modes $a_2$ and $a_3$ are entangled and the subsequent beam-splitter acts as a perfect swap between modes $a_1$ and $a_3$. Since $n_1^{\mathrm{th}} \neq 0$, the output of mode $a_3$ will receive the unwanted thermal noise while the output of mode $a_1$ forms a two-mode squeezed state with mode $a_2$. Provided that the input for mode $a_2$ is also vacuum noise, then the shared state for the output of modes $a_1$ and $a_2$ will be a two-mode squeezed vacuum state with maximum purity and entanglement, unaffected by the value of $n_1^{\mathrm{th}}$, as seen in Fig.~\ref{fig:thermal_ent_purity_phase} ~(for details of the scatttering behaviour, see Appendix~\ref{app:smatsymm}).

At this point of operation, a complementary circuit description can be obtained using the \textit{right} polar decomposition of the scattering matrix instead, which also takes on a simple form,
\begin{equation}\label{eq:PDMinusPhiAlt}
 	- \mathbf{S} = \mathbf{R}^{(1,2)} \, \mathbf{U}^{(1,3)}_{\mathrm{swap}} \text{ when } \phi = + \frac{\pi}{2}
\end{equation}
where $\mathbf{R}^{(1,2)}$ is a two-mode squeezing operation between modes $a_1$ and $a_2$: 
\begin{equation}\label{eq:PolarDecompAlt}
	\mathbf{R}^{(1,2)} \leftrightarrow \exp \left[2 i \, \artanh \Big(\!\sqrt{\mathcal{C}_{23}}\Big) \left(\hrm a_1 \hrm a_2 + \hat a_1 \hat a_2 \right) \right].
\end{equation}
Then, we can write the covariance matrix described by Eq.~(\ref{eq:VThPlus1}) in an equivalent form:
\begin{equation}\label{eq:VThPlus2}
    \mathbf{V} = \mathbf{R}^{(1,2)} \mathbf{U}^{(1,3)}_{\mathrm{swap}} \mathbf{V}_{\mathrm{in}} \big(\mathbf{U}^{(1,3)}_{\mathrm{swap}}\big)^T \! \big(\mathbf{R}^{(1,2)}\big)^T.
\end{equation}
Here, the action of the beam-splitter on $\mathbf{V}_{\rm in}$ can be seen explicitly: it swaps the thermal noise from mode $a_1$ with the input from mode $a_3$, which is vacuum noise. This is followed by a two-mode squeezer acting directly to entangle modes $a_1$ and $a_2$, creating a state with maximum purity and entanglement. It is important to note that mode $a_1$ is not cooled using this scheme, and that the nonreciprocal loop only allows for the thermal noise to be rerouted so as to not appear in the output field.

If we tune $\phi$ away from operating points of perfect nonreciprocity and impedance matching, the entanglement and purity of modes $a_1$ and $a_2$ degrade as before. However, comparing Fig.~\ref{fig:thermal_ent_purity_phase} and the previous results when only considering vacuum inputs (see Fig.~\ref{fig:vacuum_ent_purity}) we note that the degradation is more pronounced when the input to mode $a_1$ is thermal. This observation further highlights the importance of nonreciprocity in implementing perfect swaps of thermal inputs.

Finally, while thermal noise in one mode is detrimental to both entanglement and purity, the effects are compounded when both modes contain some thermal noise input. Fig.~\ref{fig:thermal_variation} demonstrates the effects of  incident thermal noise on both modes of an entangled pair. The nonreciprocal loop has the benefit that regardless of the amount of thermal noise incident on mode $a_2$, the thermal noise incident on mode $a_1$ is always swapped away in the output. The usual two-mode squeezed state, on the other hand, will experience extra degradation of the entanglement and purity as $n_1^{\mathrm{th}}$ increases for even relatively small values of $n_2^{\mathrm{th}}$. In addition, provided that the noise incident on mode $a_1$ is at a higher temperature than the noise incident on mode $a_3$, $n_1^{\rm{th}} > n_3^{\rm{th}}$, then the NRL will always improve the fidelity of the entangled output state for modes $a_1$ and $a_2$ when compared to the usual TMS at the same interaction strength. This is true even when other internal (unmonitored) loss channels are present (see Appendix~\ref{app:internal}).

\subsection{Entangling the output fields of two hot modes}

\begin{figure}[t]
		\centering
		\includegraphics[width=0.8\linewidth]{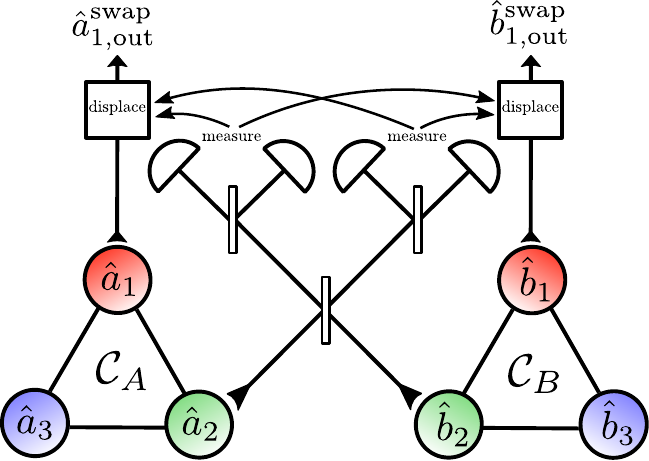}
		\caption{The schematic for a potential entanglement-swapping scheme to entangle the output fields of two hot modes. The scheme begins with two nonreciprocal loops, each comprising a hot primary mode (red), a secondary mode (green) that we wish to optimally entangle with the hot mode, and a cold auxiliary mode (blue). As with the setup covered in this paper, the hot and cold auxiliary modes are coupled via a beam-splitter interaction. The other interactions must then be two-mode squeezers. To entangle the output of both hot modes, the output state of the secondary modes is passed through a 50/50 beam-splitter and then measured. The displacement of the output for the hot modes is then conditioned on the measurements; their outputs will then be entangled.}
		\label{fig:thermal_swapping}
\end{figure}

The nonreciprocal loop allows for the output fields of one hot mode and one cold mode to be entangled with maximum purity. As seen in Fig.~\ref{fig:thermal_variation}, when using this system to realise such an entangled state between the output fields of two hot modes, the logarithmic negativity is reduced, since it is only possible to reroute the thermal noise from one of the input modes.

However, it is possible to entangle the outputs of two hot modes if we use two nonreciprocal loops and employ the entanglement-swapping protocol~\cite{hoelscher2011optimal,van2002quantum}. A possible setup is shown in Fig.~\ref{fig:thermal_swapping}, with two realizations of the NRL ($A$ and $B$, respectively). Two hot modes, $a_1$ and $b_1$, are coupled to cold auxiliary modes, $a_3$ and $b_3$, respectively, with secondary modes, $a_2$ and $b_2$, completing the corresponding loops. Both loops are operated under symmetric configurations and are thus parameterised by a single two-mode squeezing cooperativity each ($\mathcal{C}_A$ and $\mathcal{C}_B$ respectively). In light of our previous results, we also operate at points of perfect nonreciprocal scattering and impedance matching. The loop phase is assumed to be tuned such that the thermal noise incident on the hot mode is routed to the output of the cold auxiliary mode. Provided that the input for each secondary mode is also in the vacuum state, the output for the hot and secondary modes from each loop will be optimally entangled as a two-mode squeezed vacuum state with maximum purity.

In the swapping protocol, the outputs from the secondary modes are mixed via a beam-splitter and then measured. The outputs for the hot modes are then displaced, conditioned on the result of these measurements. Perfect application of this entanglement-swapping protocol would allow for the output fields of the two hot modes to be combined to produce another two-mode squeezed vacuum state with purity $\mu = 1$. The entanglement for the resulting state is then~\cite{van2002quantum}
	\begin{equation}
	E_N = 2 r \text{ where } \tanh(r) = \frac{4 \sqrt{\mathcal{C}_A \mathcal{C}_B}}{(1+\mathcal{C}_A)(1+\mathcal{C}_B)}
	\end{equation}
which is independent of the thermal noise incident on both hot modes. We therefore see that the entanglement of the propagating fields from hot modes $a_1$ and $b_1$, and the purity of the generated flying states, can be rendered robust against their thermal inputs using nonreciprocity.

\section{Conclusions and outlook}

In this paper, we analyse how engineered nonreciprocity in quantum systems influences their ability to entangle propagating fields incident on their constituent quantum modes. This requires analysing the role of nonreciprocity beyond asymmetric scattering for signal routing - a notion that can be defined completely classically - and, in particular, exploring its influence on quantum correlations and steady-state entanglement of fields. Using a minimal system consisting of a two-mode squeezer where each mode is coupled to a third auxiliary mode in a closed-loop configuration, we show that the generated entanglement of output fields depends strongly on the direction of nonreciprocal scattering. It is not \textit{a priori} obvious that it should be possible to entangle outputs from two modes for which signal flow is only unidirectional. However, we show that this is indeed possible, given the right configuration of the system.

To explain this somewhat surprising entanglement behaviour, we develop a heuristic picture that is based on a polar decomposition of the scattering matrix. This description maps nonreciprocal scattering to sequential Gaussian circuit operations, including pairwise beam splitters and, more importantly, two-mode squeezers which are necessary for generating entanglement. This picture helps us to explain a second key result: that engineered nonreciprocity can be used to reroute thermal fluctuations from a hot propagating mode toward the output of the cold auxiliary mode (via the beam-splitter component), while simultaneously allowing the entanglement of propagating output fields (via the two-mode squeezer component). This renders output-field entanglement much more robust to thermal fluctuations when compared to a reciprocal two-mode squeezing interaction.

Our work is relevant to the generation of stationary entanglement of itinerant low-frequency modes, where thermal occupations can be appreciable even at cryogenic temperatures. Our analysis also brings to light the possible uses of nonreciprocity in entanglement generation. With recent interest in multipartite entanglement in quantum systems of increasing scale, our work invites the exploration of whether engineered nonreciprocity can be a useful resource in improving robustness of multipartite entanglement in low-frequency modes.

\section{Acknowledgements}
\label{sec_thanks}
We thank C.~M. Wilson and C.~W. Sandbo Chang for the useful discussions. A.M. and L.O. acknowledge funding by the Deutsche Forschungsgemeinschaft through the project CRC 910 and the Emmy Noether program (Grant No. ME 4863/1-1). S.K. acknowledges funding by the Israeli Science Foundation (Grant No. 1794/21) and the Israeli Innovation Authority (Grant No. 73754). We acknowledge support by the KIT-Publication Fund of the Karlsruhe Institute of Technology.

\appendix

\section{Experimental implementation in parametric cQED}
\label{app:parCQED}

In this appendix section we detail how the model we consider in this work, Eq.~(\ref{eq:HNRL}), can be realised with standard techniques in parametric cQED.

\subsection{Circuit Lagrangian}

The minimal system we require consists of three modes coupled via a single nonlinear mixing element. The circuit Lagrangian for such a system is given by:
\begin{align}
    \mathbb{L} = \sum_{m=1,2,3} \! \left(  \frac{1}{2}C_m\dot{\tilde{\Phi}}_m^2 - \frac{\tilde{\Phi}_m^2}{2L_m} \right) - U_{J}(\{\tilde{\Phi}_m\})
    \label{appeq:circL}
\end{align}
where $U_{J}(\{\tilde{\Phi}_m\})$ describes the energy due to the inductance of superconducting circuit elements incorporating Josephson junctions. Expressing $U_{J}(\{\tilde{\Phi}_m\})$ using a Taylor expansion around an equilbrium point, $\{\tilde{\Phi}_m \} = 0$, gives the following form:
\begin{align}
    U_{J} = \sum_{mn} \tilde{c}^{(2)}_{mn} \tilde{\Phi}_m\tilde{\Phi}_n + \sum_{mnr}\tilde{c}^{(3)}_{mnr}\tilde{\Phi}_m\tilde{\Phi}_n\tilde{\Phi}_r + \ldots
\end{align}
where
\begin{subequations}
    \begin{align}
    \tilde{c}^{(2)}_{mn} &= \frac{1}{2!}\frac{\partial^2 U_{\rm NL}}{\partial \tilde{\Phi}_m\partial \tilde{\Phi}_n}\Bigg|_{\{\tilde{\Phi}_m\} \to 0},
     \\
    \tilde{c}^{(3)}_{mnr} &= \frac{1}{3!}\frac{\partial^3 U_{\rm NL}}{\partial \tilde{\Phi}_m\partial \tilde{\Phi}_n\partial \tilde{\Phi}_r}\Bigg|_{\{\tilde{\Phi}_m\} \to 0}.
\end{align}
\end{subequations}

The Josephson term provides quadratic contributions that serve to renormalise the bare linear modes, as we will soon show. Earlier work~\cite{lecocq2017nonreciprocal} has considered the time-modulation of these quadratic terms as a means of generating tunable Gaussian interactions. 

While this is a viable approach, in this appendix section we will instead explore means of realising the NRL by coherent pumping of nonlinear mixing terms, demonstrated in several recent works~\cite{chien_multiparametric_2020, zhou_modular_2022}. The lowest-order nonlinear contribution is defined as
\begin{align}
    U_{\rm NL} = \sum_{mnr}\tilde{c}^{(3)}_{mnr}\tilde{\Phi}_m\tilde{\Phi}_n\tilde{\Phi}_r.
    \label{appeq:unl}
\end{align}
Note that higher-order nonlinear contributions cannot always be neglected, as they can contribute non-rotating terms, such as Kerr terms. However, recent work~\cite{frattini_3-wave_2017, sivak_kerr-free_2019} has shown methods to engineer Kerr-free nonlinear potentials which can substantially suppress such contributions.

Now, by introducing the vector of mode fluxes $\bm{\tilde{\Phi}} = (\tilde{\Phi}_1,\tilde{\Phi}_2,\tilde{\Phi}_3)$, the circuit Lagrangian of Eq.~(\ref{appeq:circL}) can be written in the compact matrix form
\begin{align}
    \mathbb{L} = \frac{1}{2}\dot{\bm{\tilde{\Phi}}}{\!\!~}^T\mathbf{C}\dot{\bm{\tilde{\Phi}}} - \frac{1}{2}\bm{\tilde{\Phi}}^T\mathbf{L}^{-1}\bm{\tilde{\Phi}} - U_{\rm NL}(\{\tilde{\Phi}_m\})
    \label{appeq:linL}
\end{align}
where we have introduced the capacitance matrix $\mathbf{C}$ and inductance matrix $\mathbf{L}$, whose matrix elements are defined as:
\begin{align}
    \mathbf{C}_{nm} = C_n \delta_{nm},~   \mathbf{L}^{-1}_{nm} = L_n^{-1}\delta_{nm} + \tilde{c}^{(2)}_{nm}
\end{align}
where $\delta_{nm}$ is the Kronecker $\delta$ function. It now proves useful to diagonalise the quadratic part of the circuit Lagrangian. To do this, we first write the Euler-Lagrange equations for the quadratic part of the Lagrangian,
\begin{align}
    \mathbf{C}\ddot{\bm{\tilde{\Phi}}} = -\mathbf{L}^{-1}\tilde{\bm{\Phi}}.
\end{align}
The Euler-Lagrange equations allow us to introduce dimensionless circuit eigenmodes $\{\bm{\varphi}^{(j)}\}$ with eigenfrequencies $\{\omega_j\}$, which satisfy the generalised eigenproblem
\begin{align}
    \omega_j^2 \mathbf{C}\bm{\varphi}^{(j)} = \mathbf{L}^{-1}\bm{\varphi}^{(j)}
\end{align}
and obey the orthogonality relations
\begin{align}
    \bm{\varphi}^{(j) T} \mathbf{C} \bm{\varphi}^{(k)} = C_S \, \delta_{jk},~\bm{\varphi}^{(j) T} \mathbf{L}^{-1} \bm{\varphi}^{(k)} = C_S \, \omega_k^2 \delta_{jk}
\end{align}
where $C_S$ is a scaling capacitance introduced to ensure that the circuit eigenmodes are dimensionless. Note that $C_S \, \omega_j^2$ has units of inverse inductance, as required by the above expression. 

The eigenmodes form a complete basis, which allows us to expand the circuit flux variables in terms of flux variables corresponding to the eigenmodes,
   \begin{align}
    \bm{\tilde{\Phi}} = \sum_j \Phi_j \bm{\varphi}^{(j)},~    \dot{\bm{\tilde{\Phi}}} = \sum_j \dot{\Phi}_j \bm{\varphi}^{(j)},
\end{align} 
Substituting the above into Eq.~(\ref{appeq:linL}) and making use of the orthogonality of the circuit eigenmodes, we immediately arrive at the Lagrangian,
\begin{align}
\mathbb{L} = \frac{1}{2}\sum_j C_S \dot{\Phi}_j^2 - \frac{1}{2}\sum_j C_S \omega_j^2 \Phi_j^2 - U_{\rm NL}(\{\Phi_j\})
\end{align}
where, using Eq.~(\ref{appeq:unl}), the nonlinear contribution in terms of eigenmode fluxes takes the form:
\begin{align}
    & U_{\rm NL}(\{\Phi_j\}) = \sum_{mnr}\tilde{c}^{(3)}_{mnr} \sum_{jkl} \varphi_m^{(j)}\varphi_n^{(k)}\varphi_r^{(l)} \Phi_j\Phi_k\Phi_l \nonumber \\
    &= \sum_{jkl}\!\left[\sum_{mnr}\tilde{c}^{(3)}_{mnr}\varphi_m^{(j)}\varphi_n^{(k)}\varphi_r^{(l)}\right]\!\!\Phi_j\Phi_k\Phi_l \equiv \sum_{jkl} c'_{jkl} \Phi_j\Phi_k\Phi_l.
\end{align}

In this diagonalised form, the conjugate momenta are simply given by $\frac{\partial \mathbb{L}}{\partial \dot{\Phi}_j} = Q_j = C_S \dot{\Phi}_j$, so that the Legendre transformation defining the Hamiltonian, $\mathbb{H} = \sum_j Q_j\dot{\Phi}_j - \mathbb{L}$, may be carried out straightforwardly. We finally obtain the circuit Hamiltonian,
\begin{align}
    \mathbb{H} = \frac{1}{2}\sum_j \frac{Q_j^2}{C_S} + \frac{1}{2}\sum_j C_S \omega_j^2 \Phi_j^2 + U_{\rm NL}(\{\Phi_j\}).
\end{align}

We can now obtain the quantum Hamiltonian by promoting the canonical position and momenta $\Phi_j, Q_j$ to operators, then writing them in the basis of creation and annihilation operators satisfying the usual commutation relations $[\hat{d}_j,\hat{d}^{\dagger}_{k}] = \delta_{jk}$,
\begin{subequations}
\begin{align}
    \hat{\Phi}_j &= \sqrt{\frac{1}{2C_S \omega_j}} \left(\hat{d}_j+\hat{d}_j^{\dagger}\right) \equiv \sqrt{\frac{\hbar Z_j}{2}}\left(\hat{d}_j+\hat{d}_j^{\dagger}\right) \\
    \hat{Q}_j &= -i\sqrt{\frac{C_S \omega_j}{2}} \left(\hat{d}_j-\hat{d}_j^{\dagger}\right) \equiv -i\sqrt{\frac{\hbar}{2Z_j}}\left(\hat{d}_j-\hat{d}_j^{\dagger}\right)
\end{align}  
\end{subequations}
where $Z_j$ defines the effective impedance of the $j$th circuit eigenmode.

In terms of mode creation and annihilation operators, the system Hamiltonian then takes the form
\begin{align}
    \hat{\mathbb{H}}/\hbar = \sum_{j} \omega_j \hat{d}_j^{\dagger}\hat{d}_j + \sum_{jkl} c_{jkl} (\hat{d}_j + \hat{d}_j^{\dagger})(\hat{d}_k + \hat{d}_k^{\dagger})(\hat{d}_l + \hat{d}_l^{\dagger})
    \label{appeq:Hfull}
\end{align}
where $c_{jkl} = \sqrt{\hbar Z_jZ_kZ_l/8}~c'_{jkl}$.

Note that throughout this derivation we have made no assumptions regarding the frequencies $\{\omega_j\}$ of the modes. In particular, Eq.~(\ref{appeq:Hfull}) can describe a system with a single ``hot'' low-frequency (RF) mode coupled to two cooler higher frequency (microwave) modes, as analysed in the main text for rerouting of thermal fluctuations.

\subsection{NRL Hamiltonian using parametric drives }

We now introduce parametric drives to Eq.~(\ref{appeq:Hfull}) that allow us to realise the NRL proposed in the main text. In what follows, we set $\hbar=1$ and consider three coherent pump tones at frequencies $\{\nu_j\}$ with generally complex pump amplitudes $|\alpha_j|e^{-i\phi_{pj}}$, applied to the $j$th mode. In the steady-state, the pump tones lead to a coherent displacement of the system modes which can be accounted for via a standard (and exact) displacement transformation
\begin{align}
    \hat{d}_j = |\alpha_j|e^{-i\phi_{pj}}e^{-i\nu_{j}t} + \hat{a}_j e^{-i\phi_j},
\end{align}
where $\phi_j$ defines an arbitrary phase for the $j$th mode operator that leaves commutation relations unchanged. We simultaneously move to an interaction picture via the following unitary transformation:
\begin{align}
    \hat{\mathcal{U}} = \prod_j {\rm exp}\left[-i\omega_j \hat{a}_j^{\dagger}\hat{a}_j t \right],
\end{align}
to remove trivial evolution due to the bare mode Hamiltonians. The transformed Hamiltonian is then given by:
\begin{align}
    \hat{H}_{\rm NRL} = \hat{\mathcal{U}} \hat{\mathbb{H}} \, \hat{\mathcal{U}}^{\dagger} - i \, \dot{\hat{\mathcal{U}}} \, \hat{\mathcal{U}}^{\dagger}
\end{align}
We are interested in the quadratic terms in the transformed Hamiltonian (linear terms only lead to displacements while higher-order nonlinear terms are suppressed for strong enough pump amplitudes). These take the form:
\begin{align}
    &\hat{H}_{\rm NRL} = \sum_{jkl} 3 c_{jkl} (|\alpha_j| e^{-i\phi_{pj}}e^{-i\nu_j t}+ h.c.) \times \nonumber \\
    &(\hat{a}_k e^{-i\phi_k}e^{-i\omega_k t} + \hat{a}_k^{\dagger}e^{i\phi_k}e^{i\omega_k t})(\hat{a}_le^{-i\phi_l}e^{-i\omega_l t} + \hat{a}_l^{\dagger}e^{i\phi_l}e^{i\omega_l t})
    \label{appeq:htimedep}
\end{align}
Through an appropriate choices of pump frequencies, we can now make specific interactions in Eq.~(\ref{appeq:htimedep}) resonant in the interaction picture. To realise the NRL, we choose the following pump frequencies:
  \begin{align}
    \nu_1 = \omega_2 + \omega_3,~\nu_2 = \omega_3 - \omega_1,~\nu_3 = \omega_1 + \omega_2
\end{align}  
It is straightforward to see from Eq.~(\ref{appeq:htimedep}) that $\nu_1$ now resonantly pumps a two-mode squeezing interaction between modes $a_2$ and $a_3$, while $\nu_3$ pumps a two-mode squeezing interaction between the modes $a_2$ and $a_3$. In contrast, $\nu_2$ pumps a beam-splitter interaction between modes $a_1$ and $a_3$. All other interaction terms will be rapidly oscillating in this frame and can be neglected within the rotating wave approximation (RWA), which will be discussed later in this section.

Under this choice of pump frequencies, the system Hamiltonian takes the form:
\begin{align}
    &\hat{H}_{\rm NRL} = 3c_{123}\Big( |\alpha_3|e^{-i(\phi_{p3}-\phi_1-\phi_2)}\hat{a}_1^{\dagger}\hat{a}^{\dagger}_2  +\nonumber \\ 
    &|\alpha_2|e^{-i(\phi_{p2}-\phi_3+\phi_1)}\hat{a}_3^{\dagger}\hat{a}_1 + |\alpha_1|e^{-i(\phi_{p1}-\phi_2-\phi_3)}\hat{a}_2^{\dagger}\hat{a}_3^{\dagger} + h.c. \Big)
\end{align}
where we have retained only the resonant terms.

We can finally address the question of the phases. The gauge phases $\{\phi_j\}$ can be freely chosen to absorb dependencies on pump phases. In particular, requiring $\phi_{p3} = \phi_1 + \phi_2$, $\phi_{p2} = \phi_3 - \phi_1$ removes the phase dependence of the first two terms above, and yields the final system Hamiltonian
\begin{align}
    \hat{H}_{\rm NRL} = \left(g_{12}\hat{a}_1^{\dagger}\hat{a}^{\dagger}_2 + g_{13}\hat{a}_3^{\dagger}\hat{a}_1 + g_{23}e^{i\phi}\hat{a}_2^{\dagger}\hat{a}_3^{\dagger}\right) + h.c.
\end{align}
where:
\begin{align}
g_{12} = 3c_{123}|\alpha_3|,~g_{13} = 3c_{123}|\alpha_2|,~g_{23} = 3c_{123}|\alpha_1|
\end{align}
and the remaining loop phase is given by $\phi = \phi_{p3}+\phi_{p2}-\phi_{p1}$. In particular, the loop phase is independent of $\{\phi_j\}$ and hence cannot be ``gauged away.'' Furthermore, $\phi$ can be tuned by rotating the phase of any of the incident pump fields. We have thus obtained Eq.~(\ref{eq:HNRL}) of the main text, with interaction strengths tunable via pump amplitudes, having started from a very general but minimal three-mode circuit Lagrangian with a single three-wave nonlinear mixing element.

From the above derivation, it is clear that the validity of Eq.~(\ref{eq:HNRL}) hinges on the RWA. To verify that the RWA is valid, we consider a concrete three-mode system with frequencies $\omega_j/(2\pi) \in \{0.5,7.5,10.5\}~{\rm GHz}$; these are representative of modes realised in very recent RF quantum optics experiments, such as Ref.~\cite{rodrigues_cooling_2021}. Then, the desired pump frequencies are given by $\nu_j/(2\pi) \in \{18,10,8\}~{\rm GHz}$. Terms we have dropped in arriving at Eq.~(\ref{eq:HNRL}) include undesired beam splitter interactions pumped by difference frequencies $f_-/(2\pi) \in \{\omega_2-\omega_1,\omega_3-\omega_2\}/(2\pi) = \{7,3\}~{\rm GHz}$, and undesired amplifying interactions pumped by sum frequencies $f_+/(2\pi) \in \{\omega_1+\omega_3,2\omega_1,2\omega_2,2\omega_3\}/(2\pi) = \{11,1,15,21\}~{\rm GHz}$. For this particular choice of $\{\omega_j\}$, the desired pump frequencies are hence at least $500~{\rm Mhz}$ (and typically further) away from all undesired pumping frequencies and individual mode frequencies. As a result, all non-resonant interaction terms will be oscillating with a frequency of at least $500~{\rm MHz}$. The RWA is valid provided this oscillation frequency is much larger than the interaction strengths $g_{jk}$. Cooperativities $\mathcal{C}_{jk} \sim O(1)$ considered in the main text imply $g_{jk} \simeq \kappa_j,\kappa_k$, where $\kappa_j$ is the $j$th mode decay rate. For typical decay rates $\kappa_j\sim O(1)~{\rm MHz}$ in cQED, this means the fast rotating frequency is at least $O(100)$ times larger than the required interaction strengths. Hence the RWA can be expected to hold.

\section{Heisenberg-Langevin equations of motion}
\label{app:heis}
We work in the quadrature basis, which consists of the position and momentum quadrature operators, $\hat X_j = (\hrm a_j + \hat a_j)/\sqrt{2}$ and $\hat P_j = i (\hrm a_j - \hat a_j)/\sqrt{2}$, respectively. The canonical commutation relations have the usual form $\comm{\hat X_j}{\hat P_k} = i \delta_{jk}$. Defining the vector of quadrature operators for the three-mode system $\vec{R} = (\hat X_1,\hat P_1,\hat X_2,\hat P_2,\hat X_3,\hat P_3)$, we can write the Heisenberg-Langevin equations as the follows
	\begin{equation}\label{appeq:heis}
	\frac{d}{dt} \vec{R}(t) = \mathbf{M} \vec{R}(t) - \sqrt{\boldsymbol{\kappa}} \vec R_{\mathrm{in}}(t)
	\end{equation}
where $\mathbf{M}$ is a time independent dynamical matrix, $\boldsymbol{\kappa} = \mathrm{diag}(\kappa_1,\kappa_1,\kappa_2,\kappa_2,\kappa_3,\kappa_3)$ is a diagonal matrix of the mode damping rates, and $\vec{R}_{\mathrm{in}}$ is the vector of input noise operators in the quadrature basis. The correlators of the elements of the matrix $\vec{R}_{\mathrm{in}}$ have the following form:
	\begin{align}\label{appeq:correlators}
	&\expt{\hat X_{j,\mathrm{in}}(t) \hat X_{k,\mathrm{in}}(t')} = \delta_{jk} \left(n^{\rm th}_j + \frac{1}{2}\right) \delta(t-t') \nonumber \\
	&\expt{\hat P_{j,\mathrm{in}}(t) \hat P_{k,\mathrm{in}}(t')} = \delta_{jk} \left(n^{\rm th}_j + \frac{1}{2}\right) \delta(t-t') \nonumber \\
	&\expt{\hat X_{j,\mathrm{in}}(t) \hat P_{k,\mathrm{in}}(t')} = \delta_{jk} \frac{i}{2} \delta(t-t').
	\end{align}
These are Gaussian white noise processes, and so have a mean of zero. The dynamical matrix for the system can be written as follows:
    \begin{equation}
        \mathbf{M} = \begin{pmatrix}
                -\displaystyle{\frac{\kappa_1}{2}} \mathbf{I} & -g_{12} \mathbf{X} & g_{13} \mathbf{J} \\[0.7em]
                - g_{12} \mathbf{X} &  - \displaystyle{\frac{\kappa_2}{2}} \mathbf{I} & \substack{\displaystyle{g_{23} (\sin \phi \mathbf{Z}} \\[0.2em] \displaystyle{- \cos \phi \mathbf{X})}} & \\[1.0em]
                g_{13} \mathbf{J} & \substack{\displaystyle{g_{23} (\sin \phi \mathbf{Z}} \\[0.2em] \displaystyle{- \cos \phi \mathbf{X})}} & - \displaystyle{\frac{\kappa_3}{2}} \mathbf{I}
        \end{pmatrix}
    \end{equation}
where we have written the matrix in block form using the following $2 \times 2$ matrices
    \begin{equation}
        \mathbf{X} = \begin{pmatrix} 0 & 1 \\ 1 & 0 \end{pmatrix} \hspace{4mm} \mathbf{Z} = \begin{pmatrix} 1 & 0 \\ 0 & -1 \end{pmatrix} \hspace{4mm}
        \mathbf{J} = \begin{pmatrix} 0 & 1 \\ -1 & 0 \end{pmatrix}
        \label{appeq:XJdef}
    \end{equation}
while $\mathbf{I}$ is the $2 \times 2$ identity matrix and $\mathbf{0}$ the $2 \times 2$ zero matrix. The linear Heisenberg-Langevin equations of motion, Eq.~(\ref{appeq:heis}), can be transformed to frequency space and written in the compact form
	\begin{equation}
	-i \omega \vec R[\omega] = \mathbf{M} \vec R[\omega] - \sqrt{\boldsymbol{\kappa}} \vec R_{\mathrm{in}}[\omega].
	\end{equation}
This linear algebraic system can be straightforwardly solved,
    \begin{equation}\label{appeq:heisW}
	\vec R[\omega] = -(i\omega+\mathbf{M})^{-1} \sqrt{\boldsymbol{\kappa}} \vec R_{\mathrm{in}}[\omega].
	\end{equation}
Quantum input-output theory~\cite{gardiner_input_1985} relates the output field quadratures to the input and system fields:
    \begin{equation}
    \vec R_{\mathrm{out}}[\omega] = \vec R_{\mathrm{in}}[\omega] + \sqrt{\boldsymbol{\kappa}} \vec R[\omega].
    \label{appeq:qio}
    \end{equation}
Using Eq.~(\ref{appeq:heisW}) then allows us to express the output fields entirely in terms of the input fields, which defines the scattering matrix $\mathbf{S}[\omega]$, as introduced in Eq.~(\ref{eq:smatdef}) of the main text. The covariance matrix for the output modes in frequency space may be written as
\begin{align}
	& \mathbf{V}[\omega] = \nonumber \\
	& \frac{1}{2} \infint \!\!\! \expt{\vec R_{\mathrm{out}}[\omega] \vec R_{\mathrm{out}} [\omega']^T + \vec R_{\mathrm{out}}[\omega'] \vec R_{\mathrm{out}} [\omega]^T} d \omega'
    \label{appeq:vout}
\end{align}
where we take the outer product of the $\vec{R}_{\mathrm{out}}$ vectors. The above can be rewritten in terms of the input fields and the scattering matrix using Eq.~(\ref{eq:smatdef}), and the correlators of the input fields may be calculated using the frequency space version of the correlators from Eq.~(\ref{appeq:correlators}), where the Fourier transform replaces $\delta(t-t')$ with $\delta(\omega+\omega')$. Evaluating the integral in Eq.~(\ref{appeq:vout}) will then yield Eq.~(\ref{eq:VW}) from the main text.
    
\section{Block form of full scattering matrix}
\label{app:smatblock}

Using Eq.~(\ref{appeq:heisW}) in conjunction with Eq.~(\ref{appeq:qio}) allows us to obtain the full, frequency-dependent scattering matrix of the three-mode system. However, as discussed in the main text, we are typically interested in scattering properties at $\omega=0$. Furthermore, we have also discussed how the simplest intuitive scattering behaviour can be analysed in the symmetric, impedance-matched case defined by $\mathcal{C}_{12},\mathcal{C}_{23} \equiv \mathcal{C}$, and $\mathcal{C}_{13}=1$. In this case, $\mathbf{S}[0] \equiv \mathbf{S}$ is determined entirely by $\mathcal{C}$ and the loop phase $\phi$, and takes the form:
\begin{widetext}
\begin{align}
    \mathbf{S} = 
    \begin{pmatrix}
    \mathbf{S}_{11} & \mathbf{S}_{12} & \mathbf{S}_{13} \\
    \mathbf{S}_{21} & \mathbf{S}_{22} & \mathbf{S}_{23} \\
    \mathbf{S}_{31} & \mathbf{S}_{32} & \mathbf{S}_{33} \\
    \end{pmatrix} \equiv D(\phi)\begin{pmatrix}
    \mathbf{F}(\phi) \cos\phi & \mathbf{B}(\phi)(1+\sin\phi) & \mathbf{A}(\phi) \\
    \mathbf{B}(\phi-\pi)(1-\sin\phi) & -(1-\mathcal{C}^2)\mathbf{I}+\mathbf{\bar{F}}(\phi)\cos\phi & \mathbf{C}(\phi) \\
    \mathbf{A}(-\phi) & \mathbf{\bar{C}}(\phi) & \mathbf{F}(\phi)\cos\phi \\
    \end{pmatrix}.
    \label{appeq:S0phi}
\end{align}
\end{widetext}
Here $D(\phi) = (1-\mathcal{C})^2 + \mathcal{C}^2\cos^2\phi$ is an overall multiplicative factor that does not influence the nonreciprocity of scattering. The diagonal terms describing reflections take the form:
\begin{align}
    &\mathbf{F}(\phi) = \mathcal{C}^2\cos(\phi) \, \mathbf{I} - (1-\mathcal{C})\mathcal{C} \, \mathbf{J} \nonumber \\
    &\mathbf{\bar{F}}(\phi) = \mathcal{C}^2\cos(\phi) \, \mathbf{I} + 2\mathcal{C} \, \mathbf{J}.
\end{align}
The off-diagonal terms that describe transmission between modes $a_1$ and $a_2$ take the form:
\begin{align}
    \mathbf{B}(\phi) = \, & \sqrt{\mathcal{C}} \frac{\cos\phi(1+\mathcal{C}\sin\phi)}{1+\sin\phi} \mathbf{Z} \nonumber \\ & - \sqrt{\mathcal{C}}(2\mathcal{C}-\mathcal{C}\sin\phi-1) \, \mathbf{X}.
\end{align}
The interaction between modes $a_1$ and $a_3$ is a beam-splitter and is compactly described by a single $\phi$-dependent matrix $\mathbf{A}(\phi)$:
\begin{align}
    \mathbf{A}(\phi) = \,\, & \mathcal{C}^2\cos\phi(1+\sin\phi) \mathbf{I} \nonumber \\ 
    & +\left[-\mathcal{C}^2\cos^2\phi-(1-\mathcal{C})(1+\mathcal{C}\sin\phi)\right] \mathbf{J}.
\end{align}
Finally, the interaction between modes $a_2$ and $a_3$ is described by:
\begin{align}
    \mathbf{C}(\phi) &= -\sqrt{\mathcal{C}}(1+\sin \phi)(1-\mathcal{C}\sin \phi)\mathbf{Z} \nonumber \\
    &~~~+ \sqrt{\mathcal{C}}\cos\phi(1- 2\mathcal{C}-\mathcal{C}\sin\phi) \mathbf{X}
    \nonumber \\
    \mathbf{\bar{C}}(\phi) &= +\sqrt{\mathcal{C}}(1-\sin \phi)(1+\mathcal{C}\sin \phi)\mathbf{Z} \nonumber \\
    &~~~+ \sqrt{\mathcal{C}}\cos\phi(1- 2\mathcal{C}+\mathcal{C}\sin\phi) \mathbf{X}.
\end{align}

From Eq.~(\ref{appeq:S0phi}), it is now straightforward to read off conditions for specific desired scattering properties. For example, impedance matching of mode $a_1$ demands $||\mathbf{S}_{11}|| = 0$, which clearly requires that $\cos \phi = 0$, and hence $\phi = \pm \pi/2$. Similarly, perfect nonreciprocal scattering between modes $a_1$ and $a_2$ as defined in Eq.~(\ref{eq:nij}) of the main text, $\mathcal{N}^{(1,2)} = 1$, clearly requires that $(1\pm\sin\phi) = 0$, which again implies that $\phi=\pm \pi/2$. These are the conditions shown in Fig.~\ref{fig:circuit_diagram}.

\section{Comparison of the scattering properties of the NRL and TMS}
\label{app:smatsymm}

When the NRL is optimised to allow swapping of thermal noise from mode $a_1$ to mode $a_3$, we use the impedance-matched case from the previous section, additionally setting the phase to $\phi = \pi/2$. In the quadrature basis the steady-state scattering matrix will have the following form: 
    \begin{equation}
       \mathbf{S}_{\mathrm{NRL}} = \begin{pmatrix}
                \mathbf{0} & \displaystyle{\frac{2 \sqrt{\mathcal{C}}}{1-\mathcal{C}}} \mathbf{X} & - \displaystyle{\frac{1+\mathcal{C}}{1-\mathcal{C}}} \mathbf{J} \\[0.6em]
                \mathbf{0} &  - \displaystyle{\frac{1+\mathcal{C}}{1-\mathcal{C}}} \mathbf{I} & -\displaystyle{\frac{2 \sqrt{\mathcal{C}}}{1-\mathcal{C}}} \mathbf{Z} & \\[0.6em]
                - \mathbf{J} & \mathbf{0} & \mathbf{0}
        \end{pmatrix}.
    \end{equation}
Replacing the cooperativity with the following squeezing parameter $r = \mathrm{artanh} [2 \sqrt{\mathcal{C}}/(1+\mathcal{C})]$ we can rewrite the above scattering matrix as follows:
    \begin{equation}
        \mathbf{S}_{\mathrm{NRL}} \equiv \begin{pmatrix}
                \mathbf{0} & \sinh(r) \mathbf{X} & - \cosh(r) \mathbf{J} \\
                \mathbf{0} &  - \cosh(r) \mathbf{I} & -\sinh(r) \mathbf{Z} & \\
                - \mathbf{J} & \mathbf{0} & \mathbf{0}
        \end{pmatrix}.
    \end{equation}
This makes clear the behaviour of the system at this point of nonreciprocity. Modes $a_1$ and $a_2$ are independent of the input noise on mode $a_1$, which only appears in mode $a_3$ showing how the noise is rerouted there. Meanwhile, modes $a_2$ and $a_3$ share some squeezed correlations.

We can also calculate the steady-state scattering matrix for an open system with a TMS Hamiltonian given by $i g (\hrm a_1 \hrm a_2 - \hat a_1 \hat a_2)$ (choosing the TMS phase to be zero as in Eq.~(\ref{eq:HTMS}) is not appropriate here; this comes from the polar decomposition):
    \begin{equation}
        \mathbf{S}_{\mathrm{TMS}} = \begin{pmatrix}
             -\displaystyle{\frac{1+\mathcal{C}}{1-\mathcal{C}}} \mathbf{I} & -\displaystyle{\frac{2 \sqrt{\mathcal{C}}}{1-\mathcal{C}}} \mathbf{Z} \\[0.6em] -\displaystyle{\frac{2 \sqrt{\mathcal{C}}}{1-\mathcal{C}}} \mathbf{Z} & -\displaystyle{\frac{1+\mathcal{C}}{1-\mathcal{C}}} \mathbf{I}
        \end{pmatrix}.
    \end{equation}
Defining the squeezing in the same way we arrive at the following scattering matrix:
    \begin{equation}
        \mathbf{S}_{\mathrm{TMS}} \equiv \begin{pmatrix}
                - \cosh(r) \mathbf{I} & -\sinh(r) \mathbf{Z} \\
                -\sinh(r) \mathbf{Z} & - \cosh(r) \mathbf{I}
        \end{pmatrix}.
    \end{equation}
While the covariance matrix of the TMS and modes $a_1$ and $a_2$ in the NRL will be identical, the scattering behaviour is markedly different. This is expected given the circuit decomposition for the NRL; since the squeezing is swapped from mode $a_3$ to mode $a_1$ by a beam-splitter, the quadratures are rotated during this swap in a manner that cannot be replicated in a TMS alone.

\section{Stability conditions}

We provide below the Routh-Hurwitz stability criterion for the three-mode loop at the phases $\phi = \pm \pi/2$:
    \begin{align}
        & 0 < \kappa_1 + \kappa_2 + \kappa_3 \nonumber \\
        & 0 < 1 - \mathcal{C}_{12} + \mathcal{C}_{13} - \mathcal{C}_{23} \nonumber \\
        & 0 < 1 - \frac{\mathcal{C}_{12}}{(1+\kappa_3/\kappa_1)(1+\kappa_3/\kappa_2)} +  \frac{\mathcal{C}_{13}}{(1+\kappa_2/\kappa_1)(1+\kappa_2/\kappa_3)}  \nonumber \\
        & \hspace{8mm} - \frac{\mathcal{C}_{23}}{(1+\kappa_1/\kappa_2)(1+\kappa_1/\kappa_3)}.
    \end{align}
Away from these phases there are more conditions which must be met, and the conditions in general take on a much more complicated form. Provided the cooperativities and dissipation rates are chosen appropriately the system can be stable for all values of the loop phase.

In case we also apply the condition $\mathcal{C}_{12} = \mathcal{C}_{13} \mathcal{C}_{23}$ to make the system nonreciprocal, the second listed condition takes on a much simpler form
    \begin{equation}
        0 < (1 - \mathcal{C}_{23})(1 + \mathcal{C}_{13}).
    \end{equation}
The system is naturally stable for all choices of the beam-splitter cooperativity, and is therefore limited by the free two-mode squeezing cooperativity $\mathcal{C}_{23} < 1$ which is identical to the stability criterion for an open two-mode squeezer. The two-mode squeezing cooperativity fixed by the nonreciprocity condition, $\mathcal{C}_{12}$, can therefore grow quite large with the system remaining stable. Again, provided the dissipation rates are chosen correctly, the other stability conditions can be satisfied as well.

\section{Thermal rerouting prevents entanglement in a reciprocal system}
\label{sec:reciprocalrerouting}

In order to reroute thermal noise away from the output of mode $a_1$ in a three-mode system while still realising entanglement between the outputs of modes $a_1$ and $a_2$, it is required that the scattering of mode $a_1$ be impedance matched and that the scattering between modes $a_1$ and $a_2$ be nonreciprocal. It is simple to demonstrate that it is not possible to do both in a three-mode system where the scattering between modes $a_1$ and $a_2$ is reciprocal. We begin with a scattering matrix which can reroute the thermal excitations away from the output of mode $a_1$:
\begin{equation}\label{eq:recipsmat}
    \mathbf{S} = \begin{pmatrix}
       \mathbf{0} & \mathbf{0} & \mathbf{S}_{13} \\
        \mathbf{0} & \mathbf{S}_{22} & \mathbf{S}_{23} \\
        \mathbf{S}_{31} & \mathbf{S}_{32} & \mathbf{S}_{33}
    \end{pmatrix}.
\end{equation}
Writing the initial covariance matrix as $\mathbf{V}_{\mathrm{in}} = \mathrm{diag}(\mathbf{V}_{1,\mathrm{in}},\mathbf{V}_{2,\mathrm{in}},\mathbf{V}_{3,\mathrm{in}})$, the covariance matrix for the output of modes $a_1$ and $a_2$ is then
\begin{align}
    & \mathbf{V}^{(1,2)} = \nonumber \\
    & \hspace{2mm} \begin{pmatrix}\label{eq:recipvmat}
       \mathbf{S}_{13} \mathbf{V}_{3,\mathrm{in}} \mathbf{S}_{13}^T & \mathbf{S}_{13} \mathbf{V}_{3,\mathrm{in}} \mathbf{S}_{23}^T \\
       \mathbf{S}_{23} \mathbf{V}_{3,\mathrm{in}} \mathbf{S}_{13}^T & \mathbf{S}_{22} \mathbf{V}_{2,\mathrm{in}} \mathbf{S}_{22}^T + \mathbf{S}_{23} \mathbf{V}_{3,\mathrm{in}} \mathbf{S}_{23}^T 
    \end{pmatrix}
\end{align}
which is independent of $\mathbf{V}_{1,\mathrm{in}}$, as desired. Assuming that Eq.~(\ref{eq:recipsmat}) is a valid scattering matrix, it must be symplectic and hence satisfy the condition $\mathbf{S} \mathbf{\Omega} \mathbf{S}^T = \mathbf{\Omega}$ where $\mathbf{\Omega} = \mathrm{diag}(\mathbf{J},\mathbf{J},\mathbf{J})$ is the symplectic form. Using this it is possible to come up with conditions for the block elements of Eq.~(\ref{eq:recipsmat}). 

First, it may be determined that $\mathbf{S}_{13} \mathbf{J} \mathbf{S}_{13}^T = \mathbf{J}$ which indicates that $\mathbf{S}_{13} \in Sp(2,\mathbb{R})$. We also have to satisfy $\mathbf{S}_{13} \mathbf{J} \mathbf{S}_{23}^T = \mathbf{0}$; since $\mathbf{S}_{13} \mathbf{J} \in Sp(2,\mathbb{R})$ must be invertible it follows that $\mathbf{S}_{23}^T = (\mathbf{S}_{13} \mathbf{J})^{-1} \mathbf{0}$ and so $\mathbf{S}_{23} = \mathbf{0}$. As a consequence, the off-diagional blocks in Eq.~(\ref{eq:recipvmat}) vanish, indicating that the output of modes $a_1$ and $a_2$ are never entangled. It is therefore not possible to realise thermal noise rerouting and entanglement in a reciprocal three-mode loop.

\section{Internal losses}
\label{app:internal}

In this appendix section we analyse the role of ``internal'' loss channels that are not used to direct inputs to, or measure outputs from, the NRL, but still contribute added noise. These channels can describe unmonitored ports of the system (for example the undercoupled port of a two-sided cavity) as well as material losses. To account for these effects, we rewrite the total loss rates for the system modes as
\begin{align}
    \kappa_j = \kappa_j^e + \kappa_j^{\rm int}
\end{align}
where $\kappa_j^e$ defines the loss rate via the monitored or external loss channel, while internal losses are described by $\kappa^{\rm int}_j$. In this way, the total loss rates are the same as those used in our analysis in the main text. Thus far, our analysis has considered $\kappa_j^{\rm int} = 0$. Accounting for couplings to additional loss channels in the case of non-zero $\kappa_j^{\rm int}$, the Heisenberg-Langevin equations can be written as:
\begin{align}
\frac{d}{dt}\vec{R} = \mathbf{M}\vec{R}(t) - \sqrt{\bm{\kappa}^e}\vec{R}_{\rm in}(t) - \sqrt{\bm{\kappa}^{\rm int}}\vec{R}_{\rm int}(t)
\end{align}
where $\bm{\kappa}^e$, $\bm{\kappa}^{\rm int}$ are matrices of external and internal losses respectively, analogous to $\bm{\kappa}$ introduced earlier. The term $\vec{R}_{\rm int}(t)$ defines additional noise introduced due to internal or generally unmonitored loss channels, which we again take to be Gaussian white noise processes with temperature $n_j^{\rm th}$ for mode $a_j$. 


\begin{figure}[t!]
		\centering
		\includegraphics[width=0.9\linewidth]{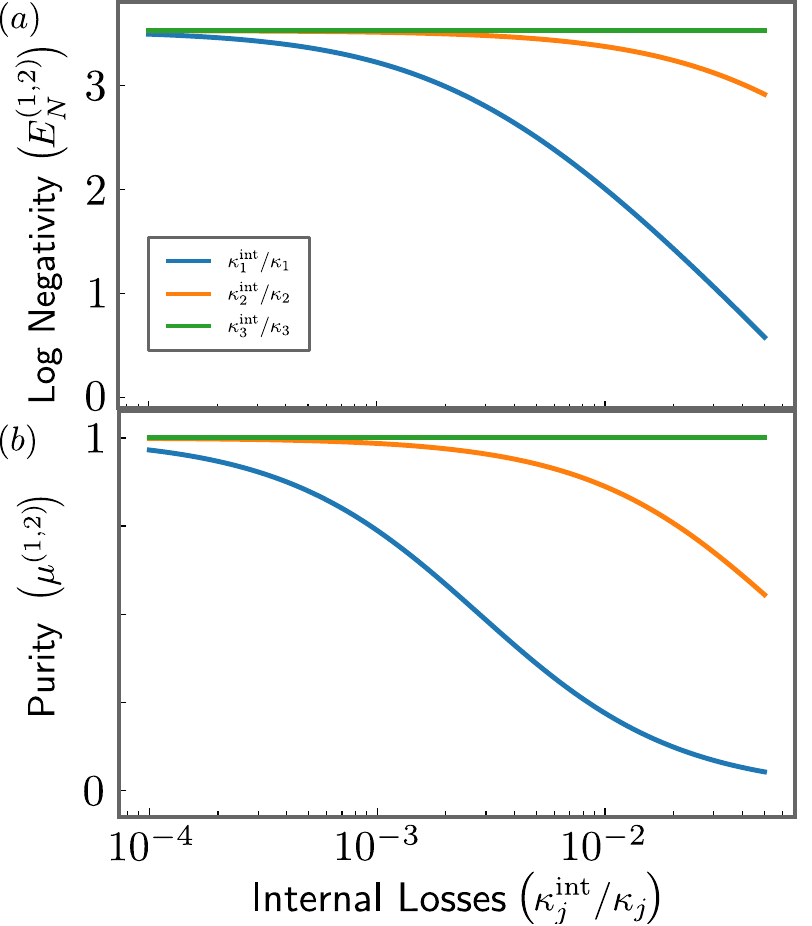}
		\caption{(a) The entanglement between the stationary output of modes $a_1$ and $a_2$ and (b) the purity of their joint quantum state as a function of the ratio of internal losses experienced by each NRL mode. We consider the symmetric configuration, with $\mathcal{C} = 0.5$, and $n^{\rm th}_1=10$, $n^{\rm th}_2=n^{\rm th}_3=0$.}
		\label{appfig:internal}
\end{figure}


Transforming to frequency space as before and rearranging, we find:
\begin{align}
    \vec{R}[\omega] = (i\omega+\mathbf{M})^{-1}\left( \sqrt{\bm{\kappa}^e}\vec{R}_{\rm in}[\omega] + \sqrt{\bm{\kappa}^{\rm int}}\vec{R}_{\rm int}[\omega]  \right).
\end{align}
Then, using the modified input-output relations for the monitored modes,
 \begin{equation}
	\vec R_{\mathrm{out}}[\omega] = \vec R_{\mathrm{in}}[\omega] + \sqrt{\boldsymbol{\kappa}^e} \vec R[\omega],
\label{appeq:qioe}
\end{equation}
we arrive at
\begin{align}
    \vec{R}_{\rm out}[\omega] &= \left[\mathbf{I}+\sqrt{\bm{\kappa}^e}(i\omega+\mathbf{M})^{-1}\sqrt{\bm{\kappa}^e}\right]\vec{R}_{\rm in}[\omega] \nonumber \\
    &~~~~+ \sqrt{\bm{\kappa}^e}(i\omega+\mathbf{M})^{-1}\sqrt{\bm{\kappa}^{\rm int}}\vec{R}_{\rm int}[\omega] \nonumber \\
    &\equiv \mathbf{S}_e[\omega]\vec{R}_{\rm in}[\omega] + \mathbf{T}_{\rm int}[\omega]\vec{R}_{\rm int}[\omega].
\end{align}
The scattering matrix for monitored channels $\mathbf{S}_e[\omega]$ is in general distinct from $\mathbf{S}[\omega]$ in Eq.~(\ref{eq:smatdef}) due to some signal being lost to unmonitored channels. Furthermore, additional noise contributions appear at monitored output ports via $\mathbf{T}_{\rm int}$ due to noise incident from these unmonitored channels. The output covariance matrix can then be calculated (once again on resonance) using Eq.~(\ref{appeq:vout}),
\begin{align}
    \mathbf{V} = \mathbf{S}_e\mathbf{V}_{\rm in}\mathbf{S}_e^T + \mathbf{T}_{\rm int}\mathbf{V}_{\rm in}\mathbf{T}_{\rm int}^T
\end{align}
which follows since $\vec{R}_{\rm in}(t)$ and $\vec{R}_{\rm int}(t)$ have no cross-correlations. 

The effect of internal losses is therefore encapsulated in the structure of the matrix $\mathbf{T}_{\rm int}$. With all pairwise interactions turned off, $\mathcal{C}_{jk}=0~\forall~j,k$, $\mathbf{T}_{\rm int}$ is a block diagonal matrix with $\mathbf{T}_{\rm int}^{(j,j)} = -2\sqrt{\kappa_j^{\rm int}\kappa_j^{e}/\kappa^2_j}~\mathbf{I}$, where $\mathbf{I}$ is the $2 \times 2$ identity matrix. Once interactions between modes are turned on to realise the NRL, the precise form of $\mathbf{T}_{\rm int}$ may be modified, but these diagonal blocks will remain non-zero. As a result, noise from unmonitored channels on mode $a_j$ will appear at the monitored port of the same mode, and can therefore effect the quantum properties of the monitored modes.

We now analyse the effect of internal losses on the NRL numerically. We will focus on modes $a_1$ and $a_2$, and therefore operate at the point of perfect nonreciprocal scattering ($\mathcal{N}^{(1,2)}=1$) that maximises entanglement of the output fields of these modes, by choosing $\phi = +\pi/2$. We also consider the symmetric NRL configuration with $\mathcal{C} = 0.5$. The resulting parameters are therefore the same as analysed in Fig.~\ref{fig:thermal_variation}.


\begin{figure}[t!]
		\centering
		\includegraphics[width=0.9\linewidth]{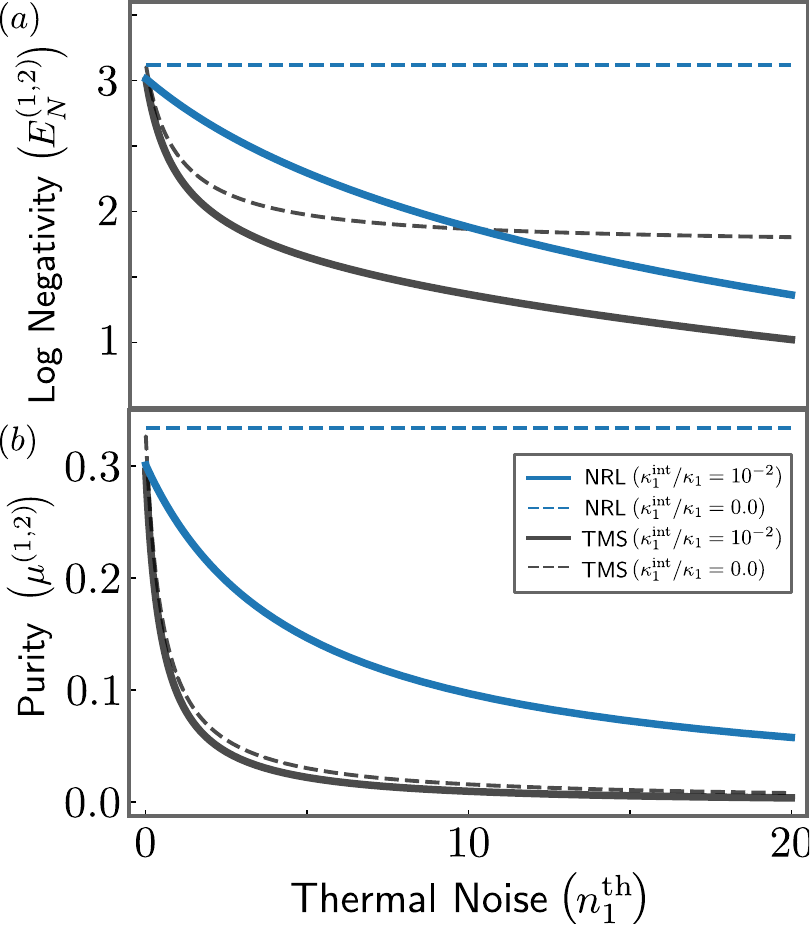}
		\caption{(a) The entanglement between the stationary output of the modes $a_1$ and $a_2$ and (b) the purity of their joint quantum state, as a function of the strength of the thermal input noise $n_1^{\mathrm{th}}$ on mode $a_1$. We compare outputs of the NRL to the TMS, both for the case of no internal losses ($\kappa_j^{\rm int}/\kappa_j = 0~\forall~j$), and for the case of a nonzero fixed internal loss ratio ($\kappa_1^{\rm int}/\kappa_1 = 10^{-2}, \kappa_2^{\rm int}/\kappa_2 = 0$). The value of $\kappa_3^{\rm int}/\kappa_3$ does not influence performance.}
		\label{appfig:internal2}
\end{figure}


Fig.~\ref{appfig:internal} shows the output field entanglement $E_N^{(1,2)}$ and purity $\mu^{(1,2)}$ for modes $a_1$ and $a_2$ as a function of the ratio of internal losses $\kappa_j^{\rm int}/\kappa_j$ for each mode. Here, $n_1^{\rm th}=10, n_{2}^{\rm th}=n_{3}^{\rm th}=0$. Note that internal losses in modes $a_1$ and $a_2$, whose entanglement is being considered, have a detrimental impact on performance. This is to be expected: noise incident on unmonitored channels for modes $a_1$ and $a_2$ appears directly, and without any quantum correlations, at the monitored ports via $\mathbf{T}_{\rm int}^{(1,1)}$ and $\mathbf{T}_{\rm int}^{(2,2)}$ respectively, which are nonzero. This uncorrelated noise reduces the fidelity and purity of the entangled state for the output of modes $a_1$ and $a_2$. Also, noise incident at a higher temperature (for the internal loss channel of mode $a_1$) is more detrimental.

However, we note that $E_N^{(1,2)}$ and $\mu^{(1,2)}$ do not depend on the internal losses of the auxiliary mode $a_3$, introduced in the NRL to enable nonreciprocal routing. There are two reasons for this effect. First, any noise from the internal loss channels of mode $a_3$ appears uncorrelated only at the monitored port of $a_3$ via $\mathbf{T}_{\rm int}^{(3,3)}$; this does not influence the output of modes $a_1$ and $a_2$. Therefore, for this noise to appear at the output of modes $a_1$ and $a_2$, it must undergo the entangling and swapping interactions of the NRL. In this process, this noise in fact seeds output field entanglement of modes $a_1$ and $a_2$, and is evidently not detrimental to the performance of the NRL.

This observation has an important implication: since only the internal losses of modes $a_1$ and $a_2$ are important, the NRL can outperform a TMS given the same internal losses and at the same interaction strength. To demonstrate this, in Fig.~\ref{appfig:internal2} we plot $E_N^{(1,2)}$ and $\mu^{(1,2)}$ for the NRL and TMS when $\mathcal{C}=0.5$, as a function of thermal input noise $n_1^{\rm th}$ with $n_2^{\rm th} = 1, n_3^{\rm th} = 0$. For simplicity, we choose only the dominant internal loss rate to be nonzero, $\kappa_1^{\rm int}/\kappa_1 = 0.01, \kappa_{2}^{\rm int}/\kappa_2 = 0$. The plotted curves have no dependence on the value of $\kappa_{3}^{\rm int}/\kappa_3$. For completeness, we also show the case of zero internal losses that was analysed in the main text. Clearly, while internal losses on the system modes are always detrimental, and have greater impact at higher thermal noise inputs, the NRL always provides higher entanglement fidelity and increased purity than the TMS. The only constraint on the NRL, then, is one we have already identified as necessary in the main text: the thermal bath fluctuations of the auxiliary mode must be cooler than those of the hot mode in the NRL, $n_3^{\rm th} < n_1^{\rm th}$.


%

\end{document}